# CCA–Fuzzy Land Cover: a new method for classifying vegetation types and coverages and its implications for deforestation analysis

H. Arellano-P.[a,b,*], J. O. Rangel-Ch.[a]

[a] *Grupo de Investigación en Biodiversidad y Conservación, Instituto de Ciencias Naturales, Universidad Nacional de Colombia, P.O. Box 7495, Bogotá D. C., Colombia*

[b] *Compensation International Progress S.A., P.O. Box 260161, Bogotá D. C., Colombia*

[*] Corresponding author at: Grupo de Investigación en Biodiversidad y Conservación, Instituto de Ciencias Naturales, Universidad Nacional de Colombia, P.O. Box 7495, Bogotá D. C., Colombia.

*E-mail address:* henryarellc@gmail.com, harellano@unal.edu.co.

ABSTRACT

Land cover has been evaluated and classified on the basis of general features using reflectance or digital levels of photographic or satellite data. One of the most common methodologies based on CORINE land cover (Coordination of Information on the Environment) data, which classifies natural cover according to a small number of categories. This method produces generalizations about the inventoried areas, resulting in the loss of important floristic and structural information about vegetation types present (such as palm groves, tall dense mangroves, and dense forests). This classification forfeits relevant information on sites with high heterogeneity and diversity. Especially in the tropics, simplification of coverage types reaches its maximum level with the use of deforestation analysis, particularly when it is reduced to the two classes of forests and nonforests. As this paper demonstrates, these results have considerable consequences for political efforts to conserve the biodiversity of megadiverse countries. We designed a new methodological approach that incorporates biological distinctiveness combined with phytosociological classification of vegetation and its relation to physical features. This approach is based on parameters obtained through canonical correspondence analysis on a fuzzy logic model, which are used to construct multiple coverage maps. This tool is useful for monitoring and analyzing vegetation dynamics, since it maintains the typological integrity of a cartographic series. The methodology creates cartographic series congruent in time and scale, can be applied to multiple and varied satellite inputs, and always evaluates the same model parameters. We tested this new method in the southwestern Colombian Caribbean region and compared our results with those from what we believe are outdated tools used in other analyses of deforestation around the world.

**1. Introduction**

In the 19th century, important theoretical contributions to biological science by E. Haeckel, T. R. Malthus, and C. Darwin led to one of its main divisions, autecology (the ecology of individual organisms or species), which subsequently influenced the Anglo-American school in the development of gradient analysis, one of two principal quantitative approaches in ecology (Jongman et al., 1995). The related biological division, synecology (the ecology of communities), was a school developed on the basis of the geography of plants and on contributions to their classification based on a physiognomic model (Humboldt, 1805, 1807). These new sources of information generated a method known today as phytosociology, which classifies vegetation according to its floristic composition. Phytosociological analysis became recognized as the second major quantitative approach in community ecology (Braun Blanquet, 1964). With the influence of these two methodologies (autecological and synecological), a close connection between vegetation and geography was established. This connection, aided by the development of statistics, produced new analytical tools that were gradually implemented in vegetation studies, notably in landscape ecology in Germany and Great Britain during the 20th century (Jongman et al., 1995).

Today, biology and ecological relationships are again a source of inspiration in creating useful complex models in different scientific areas. Along with this relatively new modeling type based on biological complexity has arisen the concept of fuzzy logic, which has been mainly associated with high-precision decision making when only parts of the processes or phenomena to be evaluated are known. The power of multivariate statistical models to synthesize, and the flexibility, versatility, and accuracy of fuzzy logic models, allow them to be powerful tools for evaluating distribution characteristics of vegetation patterns. One of the biggest obstacles encountered, however, is the lack of adequate detailed information on complex relationships between plant species and geophysical variables. Nevertheless, it is currently possible to combine valuable local information generated by vegetation analyses with geophysical variables taken from satellite data. The aim is to fill information gaps in forest coverage models, which is essential in regional development.

Notable advances in the quantification of deforestation and the gain or loss of forest coverage in the pantropics are not currently being discussed (Hansen et al., 2013; Kim et al., 2015). For this study we conducted a multivariate analysis of species distribution across environmental gradients in a region located in southwestern Colombia. Species dominance data (according to their biomass records) were combined with geographical distribution patterns of the arboreal vegetation and its many variants. Parameters of the groups and species positioned along the various environmental axes resulting from this multivariate analysis were incorporated in cartographic distribution models (fuzzy logic) for both 1987 and 2011. Comparison of the study areas during these two time periods generated deforestation measurements, as well as degradation and recovery data for the region. When we compared our results with those of the Global Forest Change (GFC) study during 2000–2013 (Hansen et al., 2013), we saw large discrepancies. This is noteworthy because deforestation and degradation data from the GFC initiative are widely accepted as accurate by governmental and academic entities worldwide. We contend that the discrepancies are due to a lack of information for evaluating the region, the need to incorporate degradation processes in the analyses, failure to recognize the main agents causing deforestation, and limited knowledge of vegetation patterns where evaluated impacts are generated. As a consequence, unsuitable methodologies are chosen (e.g., by Sánchez et al., 2012, for Colombia and by Aide et al., 2013, for Latin America and the Caribbean). The results presented here show that the total gross transformation (loss) reported for the evaluated region (gross deforestation of 18,390.85 ha year$^{-1}$ and gross degradation of 8,258.51 ha year$^{-1}$) was 26,649.36 ha year$^{-1}$, with a net loss of 10,150.56 ha year$^{-1}$. In contrast, annual gross deforestation based on data provided by Hansen et al. (2013) for the period 2000–2010 is around 9,405 ha. The pantropic results of Kim et al. (2015) show a strong correlation with Hansen et al. (2013), which thus also represent an underestimation. We are concerned about deforestation information used as baselines for initiatives such as REDD+ (Reduced Emissions from Deforestation and Forest Degradation), or in setting official deforestation values. We believe that continuing confusion in the data, especially those related to the reduction of deforestation, produce a false impression that government policies aimed at counteracting the adverse effects of climate change are on the right track. The methodology presented here produces more-accurate data for REDD+ projects and land-use planning in light of global climate change and deforestation.

## 2. Materials and methods

The study area is located in the southwestern Colombian Caribbean region, a flat area of 3,425,083 ha delimited by UTM coordinates (datum WGS 84 zone 18 N 78–72 W) 1,030,572 North, 835,192 South, 348,788 West, and 529,688 East. The analysis included regional data covering 3,297,193 ha at a scale of 1:25,000; this area included the territories of Córdoba, Sucre, Antioquia, and Bolivar departments. (Fig. 1). The area has an altitudinal variation of 0 m (sea level) to 1,839 m (measured at the source of the San Agustín River and San Juan Creek, Puerto Libertador, Department of Córdoba). The altitudinal average is 279.54 m (standard deviation [$\sigma$], 301.54; variance [$\sigma^2$], 90,929.71). The minimum slope recorded was 0º and the maximum was 59.73º (mean, 9.05º; $\sigma$, 8.56; $\sigma^2$, 73.4). The direction and aspect of topographical formations that create a rain-shadow climatic effect presented a minimum value of 0º and a maximum of 360º. The annual rainfall gradient ranges from ca. 1,343 mm in semihumid regions to over 2,890 mm in humid areas in the south and southeast.

### 2.1 Cartography information

Cartographic information was generated using traditional methods, or variations thereof, at medium resolution (scale, 1:25,000). We constructed a general description of coverage using the following variables: slope distribution, digital elevation model, direction, aspect, distribution of annual precipitation, flow and topographic convergence index (TCI). Landsat images shown in Supplementary Table 1 were used to generate mosaics that were classified using conventional methodologies and to reconstruct coverage in areas without information (i.e., areas not classifiable because of clouds and shadows). Reconstructed images were classified with a mixed methodology that included spectral signature, interactive rectification of matrices, and the visual method (Arellano & Rangel, 2010; Arellano, 2012). These methods identified coverages of large vegetal formations (s.l.) as well as forests, agricultural land, livestock areas, and scrubland.

*2.2 Regional distribution of vegetation types, using canonical correspondence analysis*

The methodology we propose is based on extensive data on vegetation, flora, and coverages with anthropic origins; on secondary information calibrated for the study region; and on physical data derived from both the digital elevation model (SRTM 30 m; USGS, 2004) and the calibrated precipitation model. Structural and floristic information was taken from Rangel et al. (2010), Estupiñan et al. (2011), Rangel & Avella (2011) and Avella & Rangel (2012), which, along with wood density information (Vásquez & Arellano, 2012), was used to develop biomass matrices. The biomass information was divided into two matrices. The first contained information from well-developed forests in superhumid to semihumid climates, including data from 62 plots, 703 species, and 14 vegetation types; the second included forests in less-humid (dry) climates, with data from 50 plots, 309 species, and 14 vegetation types. Tables 1 and 2 present information on plots and vegetation types analyzed; Fig. 2 shows the organization of vegetation types.

Vásquez and Arellano (2012), in a previous ordination analysis on part of the study area (45 plots in southern and western Córdoba), determined using detrended correspondence analysis (DCA) that the species distribution exhibited nonlinear characteristics (gradient segments). On the basis of this evidence we generated a direct gradient analysis by CCA (Canonical Correspondence Analysis), which indicated important relationships among species, plots, and environmental variables applicable to localities with conditions similar to the ones analyzed in this study.

Twenty-one physical and environmental variables were analyzed in seven groups of three in order to relate each of them to the ordination axis (three dimensions) and eliminate collinearity among variables (Arellano & Rangel, 2009, 2011). Combinations of variables that best explained data variability were (1) altitude, slope, and magnesium; (2) precipitation, iron, and sodium; and (3) altitude, pH, and phosphorus. However, because of their limited spatial representation, new variables that met the necessary requirements of the model—altitude, TCI, slope, and direction—had to be defined, as they come from different methods that we applied to the digital elevation model SRTM 30 m (1 arc seg; USGS, 2004). Annual precipitation data were also included in the analysis after calibration, filtration, and re-scaling. We can reliably arrange the data for any variable, especially when the regressions exhibit good dispersions and the eigenvalues are above 0.5 (Jongman et al., 1995).

We calculated orthogonal projections of optimal plots on each environmental gradient vector, which were re-projected on the X-, Y-, and Z-axes, respectively. To precisely define plot distributions using actual gradient values, we calibrated the ordination outcomes using multiple linear regression, based on the exclusive species' values. To build the matrix of physical and environmental variables we considered aspects such as collinearity, length of the analyzed gradient (for species and plots), relation of the data with their spatial representations, and information acquired from similar studies.

*2.3 The CCA iterative process*

The iterative process consists of repeating, for each axis, the weighted-average algorithm and data typification (standardization and normalization) until the differences between one process and the other are minimal or null (ter Braak, 1986; Jongman et al., 1995). This process consists of arranging on the first axis species and sites with maximum correlation between the weighted records (Gauch, 1982; Pielou, 1984). Values of environmental variables are analyzed using multiple regression. For subsequent axes, maximum correlation among records (which are orthogonal to one another) was again sought. Values associated with each axis represent the correlation coefficient between the weights of species and sites, with 1 (one) being the maximum value. Ter Braak (1986) and Jongman et al. (1995) discussed typification (mean 0 and variance 1), maximum correlations, regression, orthogonalization of the axes, and standardization.

In figures implementation of the CCA, plot values were extracted from species records by means of the weighted-average algorithm (WA). Another option was to derive plot values from the linear combination of environmental variables. However, in this method plot values would be predictions of a simple regression, rather than resulting from observed elements, that is, from the species compositions of the plots. This result drastically changes the location of plots in the studied gradient (Palmer, 1993). When variables are standardized, the covariance is reduced to the correlation coefficient equivalent to the scale constructed by two vectors of norm 1 (one).

When several vectors are considered, linear dependence sets the number of important variables (collinearity).

## 2.4 Orthogonalization and calibration

Because CCA results are usually given in standardized values that are difficult to read in three dimensions, we employed a new approach to analyzing vegetation: the orthogonalization and calibration of real values, which result in better visualization and incorporation into the vegetation-type models. We achieved this result by means of a multiple linear regression. Orthogonalization of plot records determined precise distances between the minimum and maximum optimums of certain vegetation types, whereas orthogonalization of species records determined the deviations from species optimums, and as well as the corresponding value of the real gradient, using exclusive species, with the goal of calibrating the remaining results to real values.

## 2.5 Basis for the classification model using fuzzy logic

Results obtained through CCA using altitudinal gradients, TCI, direction-aspect, and precipitation were the basis for construction of the potential (expected) vegetation model for the specific region. To obtain an accurate map of the distribution of true vegetation types present (whose patterns were determined by phytosociological analysis), we input information on general coverage classifications, such as primary and secondary forests and low and high shrubland (general classification or coverage). For general coverage classifications, we combined the supervised classification with visually corrected results. In addition, training surfaces (spectral seeds) were generated to reclassify pixels with similar characteristics.

We used the r.terraflow algorithm (GRASS 6.4) to calculate flow, accumulation, fill, TCI, and watersheds. The TCI describes regional patterns of soil moisture using a simple logarithmic regression between total contribution of the slopes to the specific area and the angle of each pixel with respect to the area. Sites with lower indices or less convergence are less humid than those with higher values (Lookingbill et al., 2005).

## *2.6 Reclassification of images using fuzzy logic*

Fuzzy logic makes it possible to describe complex realities, including untangling confusing signals from any classification source. The process involves constructing membership functions that explain the statistical distribution of data in the sampled space, as well as their probability of "occurrence" in a frequency range between 0 and 1. One of the great strengths of this method is the complete lack of detailed mathematical equations; instead, curves are described using data such as limits, means, standard deviations, or simply knowledge acquired in the field about potential species distributions. Data from input signals, such as those corresponding to the classes of supervised classification, generate a fuzzy set when related to the membership functions. The signal (datum) acquires a value of 1 (one) when related 100% with a function that describes certain behavior, and 0 (zero) in the opposite case. Values between 0 and 1 are also allowed and define the degree of membership of the values to one or another class of fuzzy shape.

Using this technique, we compiled detailed information based on plots, knowledge acquired in the field, and secondary sources. This increased our understanding of transition zones because, unlike other regions (e.g., the páramo), the forested areas are easily observed.  We used this method to accomplish multiple objectives, including distinguishing vegetation and coverage patterns objectively in apparently homogenous sites or in those with similar radiometric characteristics, as in the cases of Landsat coverages previously classified by traditional methods. The application of fuzzy logic is also useful in solving resolution problems when a very detailed sample exists (such as in Quickbird, WorldView-2, or IKONOS images) and information for surrounding regions needs to be amplified.

For the construction of fuzzy datasets, we used trapezoidal membership functions (trapmf) to describe the behavior of the classes derived from supervised classification with visual correction. Because of the mixture of signals and the classification of shadows and light, some resulting classes were synthesized in the same membership function; for example, the different signals that identify the forest itself were grouped into a single function, with the intention of separating major constituents by evaluating other model inputs. Gaussian curve model 2 (gauss2mf) explained the membership functions of the altitudinal signal used to describe the distribution of vegetation types along the altitudinal gradient. This step allowed us to generate greater amplitude in the signal response function and an adequate representation of vegetation limits. The behavior of vegetation in some places with steep escarpment (slopes) or with some degree of incline, as well as the description of their relationship with the direction-aspect, the TCI, and precipitation, were modeled with functions (gauss2mf).

*2.7 Coverage-model validation*

The error measurement in the model defines the level of imprecision in field records, as well as the inaccuracy of cartographic results. The ordination method using CCA is very useful in detecting known position error in the records, as it is very sensitive to regional inconsistencies in the environmental gradients evaluated. In other words, the wrong position of a plot in the sampled space usually produces ambiguous results for the position of the species records and consequently of the plots on the environmental gradients of the analysis. Because coverage models and vegetation types contain information that is in most cases continuous and transitional, the error calculation and consequently their validation can be difficult. To partially solve this difficulty, we stress the importance of specific points of observation and their interpretation, including those reported by other researchers.

After ensuring quality control of inputs and minimization of precision errors, we subjected the results obtained through modeling using fuzzy logic to tests of concordance using Cohen's kappa coefficient (Cohen, 1960; Fleiss et al., 1969; Foote & Huebner, 1995) and of receiver operating characterics (ROC), which measure the quality of the model (Slaby, 2007, Cardillo, 2008).

**3. Results**

Because of the particular climatic conditions of the study region, plot records were explored according to data on climatic affinity (to humid and dry climates), altitudinal gradient, and classification of vegetation reported for the following forest types in the cited studies: the southern humid forests of Córdoba (Avella & Rangel, 2012; Arellano & Vásquez, 2012), the flooded forests of Paramillo National Park (Estupiñan et al., 2011), the *Quercus humboldtii* forests of the Colombian Caribbean region (Rangel & Avella, 2011), and the humid and dry forests surrounding the Caribbean wetland complexes (Rangel et al., 2010). Weighted averages of species rather than linear combinations of environmental variables were used to determine optimum positions of plots with respect to the environmental gradients, in order to generate figures and their respective values in three-dimensional space. To assess results of this analysis, the Monte Carlo test was performed with about 1,000 iterations to eliminate the relationship between species and their respective plots by the random location of species records within them.

The best results in the study for the three axes evaluated were obtained using a combination of the variables precipitation, altitude, and slope for tropical forests in superhumid to semihumid climates, with the inclusion of a group in a transitional humid-to-dry climate. Location corrections were made by approximation and calibration. Centered values of relative biomass of 703 species grouped in 62 plots for these formations had a total variance of 31.37. They combined the highest eigenvalues with low probabilities of committing a type I error with regard to the null hypotheses $H_0$ (1) and $H_0$ (2), according to the random tests.

The eigenvalue for the first ordination axis was 0.907 and the maximum value of the random simulations was 0.693, with a probability of **p** = 0.001; therefore $H_0$ (1) was rejected. The explained variance was 2.9%. The Pearson correlation between matrices was 0.987. According to the Monte Carlo test, the maximum value of the different iterations of the model was 0.986, **p** = 0.001, meaning that $H_0$ (2) was rejected. The variable correlated with this axis was precipitation, and it was chosen for the final orthogonalization and calibration. For axis 2, the eigenvalue was 0.829, and the random simulation registered a maximum eigenvalue of 0.601; thus $H_0$ (1) was rejected. The explained variance was 2.6%, and the cumulative variance was 5.5%. The Pearson correlation between matrices was 0.977, and the simulated ordinations using the Monte Carlo test showed a maximum correlation of 0.969; thus $H_0$ (2) was rejected. The variable correlated with the axis was altitude, and it was chosen for orthogonalization and calibration.

For axis 3, the registered eigenvalue was 0.59, and the maximum value of the random simulations was 0.545; thus $H_0$ (1) was rejected. The variance explained by the axis was 1.9%, and the final cumulative explained variance was 7.4%. The Pearson correlation between the species and the environmental variables was 0.922 for this axis, and according to the Monte Carlo test the maximum value of the various iterations of the model was 0.962; thus in this case $H_0$ (2) was not rejected. However, the axis was analyzed with slope as a correlated variable because it represented the best result. Complete results are presented in Table 1 (3), with the chosen combinations for orthogonal calibration highlighted in gray.

Fig. 3 (A, B) shows the final distributions along precipitation, altitude, and slope gradients for vegetation patterns in superhumid to semihumid climates. Regarding the organization of plots in the multivariate space, the formation of sets closely matched results obtained in published classifications of vegetation types in humid forests of southern Córdoba department (Arellano & Vásquez, 2011; Avella & Rangel, 2012), flooded forests of Paramillo National Park (Estupiñan et al., 2011), and *Quercus humboldtii* forests in the Colombian Caribbean region. At the alliance level, species were clearly differentiated into four main groups corresponding to the following zones: elevated areas with high amounts of precipitation, mid-altitude areas with high amounts of precipitation, mid-altitude areas with intermediate amounts of precipitation, and low areas with low amounts of precipitation (Table 1). At the association level, the clusters were more dispersed, as is the case in nature. The only group showing apparently high plasticity along the gradients was the association Jacarando copaiae–Pouterietum multiflorae (BHtf2/Jco-Pmu) in plots SC-7 and SC-8. This plasticity may be due to the fact that several taxa in this community had pioneer features (i.e., faster growth, bigger leaves, light wood) (Aymard, 2015, pers. comm.)

The redundancy and dominance of *Dipterix oleifera, Cariniana pyriformis, Cavallinesia platinifolia, Jacaranda copaia, Astronium graveolens, Astrocaryum malybo, Castilla elastica*, and *Dialium guianense*, among others, are evidence of the relationship between the Cariniano pyriformis–Pentaplarietum doroteae (Bmhtf2/Cpy-Pdo) and Protio aracouchini–Viroletum elongatae (BHmhtf1/Par-Vel) associations. The vegetation types with lower plasticity occur along all gradients, although with higher concentrations at lower altitudes and in areas with lower precipitation. CCA was used to explore the shape that a vegetation pattern acquires in the distribution of species' abundance and their corresponding plots (biomass) in a given territory.

To achieve the results in the actual gradient values, CCA weights were orthogonalized (Supplementary Fig. 1 A and B). A Multiple linear regressions were used to allocate the optimal CCA values projected for the plots over the actual gradients. Fig. 4 (A, B, Table 1) shows the orthogonal projection and the final dispersion along the precipitation gradient in real values. Annual precipitation lies between 1,302.5 and 2,976.08 mm (length 1,673.58 mm). Orthogonal projections for other variables are shown in Supplementary Figures 2 and 3 (table 1). Table 2. shows the real values of the optimal vegetation types in dry climates as well as the standard deviation of their characteristic species.

*3.1 Image-classification model using fuzzy logic vegetation and other coverages*

The model of precipitation distribution and plot locations can be observed in Supplementary Fig. 4. To construct the model we incorporated the digital elevation model, using the v.vol.rst algorithm of GRASS 6.4, to enhance phenomena like rain shadows. Other cartographic inputs used in this modeling process are shown in Supplementary Figs. 5 to 10. Coverage-classification models that incorporated information from the 1980s and the first decade of the 21st century used distribution curves that were obtained from ordination and were structured in six entries, with 94 membership functions and 64 possible answers.

The resulting values were rounded to the lower limit mainly because of the design of the response set in the model. The mountain-range intervals not sampled (see ordination results, Supplementary Fig. 2A, B) were labeled according to the most probable vegetation type. Within the altitudinal range 420–610 m were located primary tropical forests of the Eschweilero antioquensis–Copaiferetum camibar association, established in superhumid climates with little or no intervention; within the range 816–1002 m were found forests of the Tovomito weddellianae–Quercetum humboldtii association.

Overall structure of the fuzzy logic model for the decades analyzed, developed using MATLAB software, is shown in Fig. 5 (data within the membership function). A preliminary supervised classification of coverages was used (see Fig. 5[1]), with five classes represented with the trapezoidal form membership functions (trapmf). The solution of the model involved the MAMDANI logic type and the MOM method for reconversion of data belonging to the fuzzy set. Model results are seen in Fig. 6 (A, B) and in Tables 4 and 5, where surface calculations for each of the determined coverages in 1987 and 2010 periods are presented.

*3.2 Coverage-model validation*

Cohen's kappa analysis of reliability and credibility and ROC quality analysis showed that coverage maps for the region constructed by fuzzy logic methodology were of high accuracy and quality, is shown in Fig. 7. The kappa value was 0.95, and the area under the ROC curve was 0.88. Because the fuzzy logic process is based on CCA ordination records that estimate optimal values for the plots, based on all species' abundance numbers, and because the method for solving fuzzy sets selects the maximum probability of occurrence of a determined coverage limited by the optimal exclusive and elective characteristic species, one would expect a significant amount of confusion in the classification of transitional or border zones and of smaller fragments, but greater accuracy for vegetation types with wide distribution and little fragmentation, primarily as a result of the effects of the probability distribution.

**4. Discussion**

The fuzzy method allowed us to find and fix inconsistencies in the geographical location of the plots, especially when information from other researchers was used. For example, the locations of plots PNP-4, PNP-6, and PNP-8 published by Estupiñan et al. (2011) included inconsistencies in the seconds reported, which were recalculated to obtain correct values. It is important to calibrate information on regional locations of plots when necessary. Otherwise, the lowest values of orthogonalization data for plots and species may be negative for a given gradient, indicating problems and inconsistencies in the location of some of the inputs used in the ordination.

Additionally, during the process of data exploration analysis, we noted that it was necessary to try various arrangements of information to find the ordination structure that represented the study area with greatest precision. This precision was achieved when the eigenvalues were increased in the evaluated matrices, considerably reducing the probability of committing type I error and of obtaining negative values in the records of calibrated gradients (with real values expressed). This study concluded that CCA, applied to the separation between gradients (i.e., humid and dry) and to the group of all plots, did not respond better to the statistical treatment applied. It is important to mention that for the group of all plots, the highest value was recorded for a type I error for $H_0$ (2) ($p = 0.99$) and for the combination of the variables slope, TCI, and direction. This was due to the lack of transitional groups; this lack, along with data located at the extremes of the gradients, limited the dispersion of data along the axes, causing a decrease in values in the matrices as well as an increased probability of committing a type I error.

In general, results showed that the variables precipitation, altitude, and TCI matched well with observations in the field, whereas slope and direction generated some questions, which were taken into account in the modeling processes. The standard deviation used was that of the characteristic, exclusive, and elective species according to the phytosociological method.

## 4.1 Comparing gross historical deforestation and coverage degradation for the region between 1987 and 2010, using data from Hansen et al. (2013)

Results of this study were compared with those reported by Hansen et al. (2013) with regard to gross deforestation (i.e., not the balance among losses, gains, and transformations, or net deforestation, but the phenomenon itself) in the study area between 2000 and 2013. Although the time periods measured are slightly different in the two studies, deforestation and gross gains between 2000 and 2010 were evaluated by both; the present study lacks information for only the three years between 2010 and 2013.

Hansen et al. (2013) recorded 99,065.52 ha deforested between 2000 and 2013. Of them, 35,878.28 ha (36.21%) coincide with the total deforested surface recorded in the present study for the period between 1987 and 2010 (recovered surfaces not included), and 19,108.36 ha were deforested since 1987 (19.28%).

Comparing the deforested surface between 2000 and 2010 reported by Hansen et al. (2013), we concluded that just 5,014.24 ha were deforested between 2010 and 2013, or 1,671.41 ha year$^{-1}$. Annual deforestation based on Hansen et al. (2013) data for the period 2000–2010 was around 9,405 ha.

These results contrast with those obtained using our methodology, which show that 165,750.20 ha were classified as having some form of deforestation by 1987, implying that for the period 1987–2010 (23 years) there was an increase in gross deforestation for forests and shrubland of 259,094.48 ha, or 11,264.97 ha year$^{-1}$. In total, in both scenarios analyzed, 424,844.68 ha were classified as deforested, and 668,885.52 ha were deforested since the beginning of the evaluated period, for an accumulated value of 927,980 ha deforested by 2010.

Coverage classified as grasslands near transitional marshes and permanent bodies of water with dispersed shrubland belonging to the *Montrichardia arborescens* community (known as "Arracachales"), which belongs to a class with elements of anthropic intervention, totaled approximately 70,777.88 ha in 1987. This implies that between 1987 and 2010, this vegetation type increased about 163,116.72 ha, at a rate of 7,092.03 ha year$^{-1}$. In total, both periods registered 233,894.60 ha, of which 214,398.68 ha have retained this vegetation pattern since 1987. For the year 2010, a total accumulation of 377,515.40 ha was recorded, and a fluctuation area was defined between gross deforestation, and coverages classified under this category ("Arracachales"), which total about 778.32 ha. A total of 466.08 ha classified in 1987 as having some degree of deforestation was converted into grasslands, and 312.24 ha of grasslands were transformed into completely deforested areas. Transitional marshes with deforested areas are difficult to work with, due to the scale, the satellite input, and characteristics of the transitional marshes themselves that make them very dynamic (the coverage moves easily), because of climatic conditions throughout the year and physical characteristics of the soil.

Hansen et al. (2013) do not take into account the degradation of forest shrubland, an important process with a very high probability of complete deforestation. We estimated 189,945.84 ha involved in this process in 23 years, with a mean of 8,258.51 ha year$^{-1}$. We therefore conclude, without taking into account the lack of coherence between given patches under this category for the region, that the estimated deforestation number provided by Hansen et al. (2013) for the region of 9,405 ha year$^{-1}$ falls far short of the total gross transformation (deforestation and degradation) reported in this study of 26,649.36 ha year$^{-1}$. Moreover, when we compared the regions of overlap of the two methods, we observed a large amount of confusion in the classification of shrubland, grasslands, and areas subject to flooding, which puts in doubt the ranges for the region presented by Hansen et al. (2013) because of the lack of verification in the field. This does not imply that the study is not valuable; it accurately notes many of the areas with no vegetation and provides an annual report of what occurred during the period evaluated. However, it is important to be cautious about using Hansen et al.'s (2013) information as a baseline for initiatives such as REDD+, or for establishing official deforestation numbers, because of the lack of data to validate information on regions evaluated, degradation of woodlands and its causes, speed and acceleration of transformations, and, above all, vegetation patterns where impacts were generated. For a full picture of deforestation in any region it is essential to perform a combined analysis that includes the phenomena of degradation, transformation, and gain, together with knowledge of vegetation types and coverages beyond a simple interpretation of forest vs. nonforest. As a result, separate analyses of individual variables involved in deforestation lead to deficient interpretations.

*4.2 Net historical deforestation and degradation between 1987 and 2010*

During the period evaluated (23 years), the region suffered a net loss (balance among deforestation, degradation, and recuperation) of its natural coverage of close to 233,463 ha (Table 5), which is the sum of the total net losses of forest-dominated surfaces (conserved forests with various levels of intervention), or 169,425 ha, and the net losses for tall and medium shrubland, around 64,038 ha. The surface net gains measured 47,780 ha for the natural vegetation. These gains can be considered net degradation when they represent an increase in shrubby communities or an increase in the structural quality of individuals (recuperation of cleared areas) for shrubland far away from the intervention, and belonging to conserved vegetation types.

Areas dominated by deforestation from anthrogenic sources registered a net addition of 185,683 ha, representing the coverage with the greatest growth in the study area. In some of the conserved forests located in foothills near Paramillo National Park, net losses by deforestation and degradation could be considered small (fewer than 5,000 ha) if compared with transformations that have occurred in the low and flat areas of the region. It is important to note that the flooding caused by the construction of the Urrá dam during this period was not taken into account in the calculations. A land-use change of this size would generate a very high transition rate, which introduces significant error in the results, their interpretations, and potential predictive models.

There are two mechanisms that result in deforestation: one involves the clearance of elements of all forest strata through logging, the other a slow process of continuous and selective degradation of some of those elements. As shown in Table 5 and Fig. 8, vegetation dominated by conserved forests presented a net loss of close to 6,900 ha, of which 4,511 ha was primary tropical forest with little or no intervention belonging to the association Protio aracouchini–Viroletum elongatae (3. BHmhtf1/Par-Vel) in very humid to superhumid climates. Another 1,309 ha of the net loss was primary tropical forest with low to intermediate intervention belonging to the association Jacarando copaiae–Pouterietum multiflorae (1. BHtf2/Jco-Pmu) in superhumid climates; and around 1,121 ha was flooded, primary tropical forest with little or no intervention belonging to the association Prestoeo decurrentis–Trichillietum poeppigi (6. Bhin1/Pde-Tpo) in superhumid climates.

As far as net gains, conserved coverages showed a minor increase of about 2,118 ha, of which 1,034 ha corresponded to primary tropical forests in superhumid climates in areas associated with water bodies and with little or no intervention, mainly of the association Macrolobio ischnocalycis–Peltogynetum purpurea (7. Bhri1/Mis-Ppu). Other components of the net gains were 622 ha of primary tropical forest with intermediate intervention belonging to the association Marilo laxiflorae–Pentaclethretum macrolobae (2. Bhmhtf2/Mla-Pma) in superhumid to very humid climates; 434 ha of primary tropical forest with little or no intervention belonging to the association Tovomito weddellianae–Quercetum humboldtii (5. Bhtf1/Twe-Qhu) in superhumid climates; and 26 ha of forests with little or no intervention belonging to the association Eschweilero antioquensis–Copaiferetum camibar (4. Bhtf1/Ean-Cca).

In forests that registered different intervention levels, the net surface losses were close to 162,483 ha (Table 5, Fig. 8), and within this classification, tropical forests with high intervention in semihumid climates belonging to the community of *Acalypha diversifolia* and *Guazuma ulmifolia* (14. Bttf3/Asp-Gul) and tropical forests in humid climates with high intervention belonging to the association Sabali mauritiiformis–Cavanillesietum platanifoliae (15. Bhtf2/Sma-Cpl) registered net losses around 10,000 ha. The maximum losses in vegetation patterns with multistratified elements were recorded for tropical forests in semihumid to humid climates with strong processes of intervention belonging to the association Cinnamomo triplinervis–Apeibetum asperae (16. Bhttf2/Ctr-Aas), with 60,309 ha, and for tropical forests in semihumid climates with high intervention belonging to the association Cordio alliodorae–Attaleetum butyraceae (23. Bttf2/Call-Abu), with around 25,574 ha lost.

The only vegetation type that showed significant net gains in surface, with an increase of 6,814 ha, was the association Xylopio aromaticae–Tapiriretum guianensis (20. Bmhtf2/Xar-Tgu), a community established in very humid climates with high intervention.

With regard to the net degradation of forests with coverages of less complexity (tall and medium shrubs), the largest increase, around 22,763 ha, was recorded for tall shrubland in semihumid to humid climates with high intervention belonging to the association Cinnamomo triplinervis–Apeibetum asperae (44. Mhttf2/Ctr-Aas), followed by an increase of about 10,000 ha for tall shrubland in semihumid climates with high intervention belonging to the association Cordio alliodorae–Attaleetum butyraceae (51. Mttf2/Call-Abu).

### 4.3 Results in global and local contexts

In recent years, restoring natural coverages in the tropics during the period 2000–2013 has been discussed extensively in specialized and popular media. The United Nations Food and Agriculture Organization (FAO) in 2010 estimated a reduction of deforestation in the tropics of around 25% between 1990 and 2010. For Colombia, the organization reported a constant number of 101,000 ha year$^{-1}$ during the same period (FAO, 2010). Official reports provided by the Institute of Hydrology, Meteorology, and Environmental Studies (IDEAM, for its initials in Spanish) noted a drastic decrease in deforestation from 310,346 ha year$^{-1}$ in 2012 (Caracol radio, 03-25-2013) to 120,933 ha year$^{-1}$ as consolidated for 2013, a reduction of around 60% (IDEAM, 2014). Although it is impossible to track the methodology producing the results of these institutions, studies such as those of Kim et al. (2015), with an analysis of several tropical countries, and Sánchez et al. (2012), addressing Colombia, give us some insights.

Following the same approach as Hansen et al. (2013), Kim et al. (2015) analyzed the phenomenon of deforestation over 20 years (1990–2010) in 34 tropical countries using a high-resolution (30-m) satellite series based on Landsat images and GLS (Global Land Survey) data. Their results support our evaluation and analysis of deforestation as far as the impossibility that deforestation numbers could be decreasing, as reported by the FAO and the IDEAM, given current worldwide conditions, including the rate of population growth and the resulting unstoppable demand for goods and services.

In addition, Kim et al. (2015) estimated an acceleration of deforestation for the pantropics of around 62% during the study period (from 4.04 x $10^6$ ha year$^{-1}$ for the decade after 1990 to 6.54 x $10^6$ ha year$^{-1}$ for the decade after 2000). However, it is important to note that this approach doesn't take into account structural degradation from multistratified arboreal vegetation to communities of less complexity, such as shrubland, an issue that significantly affects the final values as is demonstrated in this contribution. In addition, their method lacks information on vegetation patterns that dominate a given surface, which precludes explaining and contextualizing much of the transformation that occurred within the evaluated scenarios for a particular region and its surroundings.

Kim et al. (2015) concluded that their estimates for the 34 countries show a strong correlation with Hansen et al. (2013) outcomes, and that the national estimates of forest change post-2000 were correlated, with a coefficient of determination $r^2 > 0.95$. These results should be verified, since we conclude, after comparing these studies with the data presented here, that these publications significantly underestimate deforestation rates in the pantropic region. For Colombia, Kim et al. (2015) showed that, between 1990 and 2000, net deforestation reached 130,000 ha year$^{-1}$, whereas during the period 2000–2010 it reached 363,000 ha year$^{-1}$ (similar to the number reported by the IDEAM in Colombia of 310,346 ha for 2012). Nevertheless, is important to note that, according to these authors, during 2005–2010 there was a gross loss of 498,870 ha year$^{-1}$ with a gain of 54,000 ha year$^{-1}$, for a net rate of deforestation of 444,870 ha year$^{-1}$. These results were calculated without taking into account the composition and degradation of the vegetation pattern. Moreover, Kim et al. (2015) recognized that the inaccuracy and incoherence in the definition of forest accounts in part for the large discrepancies between their numbers and those reported by entities such as the FAO for the periods evaluated.

We demonstrate in the present contribution that combining multistratified arboreal vegetation with less-structured vegetation into a single class (forest), in order to compare it with areas devoid of forest and thereby define deforestation, represents a large conceptual error. This error compounds that committed by choosing an inadequate method for comparing and classifying the phenomenon. Kim et al. (2015) demonstrated that classification and comparison methods based on sampling satellite inputs at different scales are several times less precise than methods using total comparison of information at the pixel-to-pixel or wall-to-wall level. A good example of errors due to the inadequate use of methodologies (leading to confusing interpretations) is found in the work of Sánchez et al. (2012) in Colombia. In this publication a new method is presented that, according to the authors, allows the quantification of soil change at multiple spatial scales. The process consists of sampling satellite inputs and making multitemporal visual classifications between 2001 and 2010 using seven classes: woody (woody vegetation including trees and shrubs), herb (herbaceous vegetation), ag (agriculture), plant (plantations), built (urbanized areas), areas without vegetation, and water. This methodology employed more than 10,000 high-resolution QuickBird images from Google Earth, which were used as a reference for the training of the classification of MODIS multitemporal images, with a pixel resolution of 250 x 250 m, with information on the vegetation index using the statistical technique Random Forest (Sánchez et al., 2012).

According to the authors, the use of this methodology is justified because of the high values obtained in error-evaluation tests with ''out-of-bag'' (OOB) samples, which reflect a general precision of 87.4% (±4.3%). These results certainly show the statistical virtues of the method; however, they do not indicate anything about its precision and accuracy with regard to the reality of the territory, despite the existence of a coherent classification in the high-resolution QuickBird images, primarily because the MODIS sensor will never acquire spatial resolution with the applied method.

Another issue we observed in the use of this methodology by Sánchez et al. (2012) is that the attribute of type (seven classes) was assigned through a Random Forest classification using pixels with a size of 6.25 ha. This is problematic, as a study unit of this size could contain any number of coverage types, especially at the edges and in transition zones between types. For this reason, MODIS images are very useful in locating and monitoring coverage changes at the country level, because of their great temporal resolution (regularity). However, they are not appropriate for quantifying deforestation or any other phenomenon that requires a precise measurement.

For a better understanding of the preceding discussion, we can consider that Colombia has close to 114,174,800 ha, which using MODIS images would represent close to 18,267,968 pixels of 6.25 ha each. Alternatively, Landsat images would represent this land area with close to 1,268,608,888 pixels of 0.09 ha each. Therefore, the MODIS system represents in spatial information barely 1.44% of a Landsat matrix (layer, band) and is thus inadequate for evaluating the deforestation phenomenon. It is important to note that, as in the work of Hansen et al. (2013) and Kim et al. (2015), it is impossible in Sánchez et al. (2012) to differentiate the phenomena of degradation, although in this latter case it is clear that because of the large size of the pixels in the model, differentiating degradation is meaningless.

However, it is important to recognize the usefulness of Sánchez et al.'s (2012) extensive classifications of QuickBird images in the complex processes of validating coverage classifications at medium- and high-resolution scales.

These authors asserted that between 2001 and 2010, the total net increase in woody vegetation in the country approached 1,696,300 ha (188,477.77 ha year$^{-1}$), and the total net deforestation was 1,464,400 ha (162,711.11 ha year$^{-1}$). A comprehensive study of this data showed some confusion with regard to the recovery numbers. Their data indicated that the transition of totally deforested areas into regions with forests or shrubland at various levels of complexity would have been reached in just 9 years. This result cannot be explained with an argument based on the recovery of forests and shrubby vegetation with different levels of intervention, because in this study forests and shrubland were not evaluated under any of the intervention classes. The inconsistencies of the method applied by Sánchez et al. (2012) are also seen in the results of Kim et al. (2015), especially as was mentioned with regard to data for the period of 2005–2010.

The estimate for loss of primary forests in Colombia throughout its entire history is about 354,605 km$^2$; that is, 31% of the original forest has already been lost (Rangel, 2014). Rates of deforestation for the country indicate that, between 1940 and 1995, the net rate was 545,455 ha year$^{-1}$; between 1960 and 2000, 500,000 ha year$^{-1}$; and between 1970 and 2000, 400,000 ha year$^{-1}$ (Rangel, 2014). For the Middle Magdalena region (the localities of Bolívar, Cesar, Santander, Boyacá, and Cundinamarca), the estimate of annual forest loss is 81,000 ha (Cárdenas, 2006). For areas in the Caquetá department in the Amazon, the deforestation rate fluctuated between 25,000 and 40,000 ha year$^{-1}$ (Etter et al., 2006). An extreme case is the continuing deforestation in the Napo province in Ecuador, which represents for Colombia an example of severe transformation of the Amazon and reflects the effects of a chaotic process of colonization. In other parts of the Colombian Amazon, their natural isolation, particularly of the central, eastern, and western regions, has been decisive in maintaining huge areas of continuous forests in well-conserved condition.

The deforestation rate in territories protected as forest reserves (created by Colombian Law 2 of 1959) until the year 2006 was 40,000 ha year$^{-1}$. For the southwestern region of the Caribbean (SWC), between 1987 and 2010 the net annual deforestation rate was 10,150 ha year$^{-1}$ (results of this study). In 23 years the SWC lost 169,425 ha of natural forests, of which 6,900 belonged to formations with excellent ecosystem health, and 64,038 ha of tall and medium shrubland. Degraded regions increased about 47,780 ha. The gross transformation rate (deforestation and degradation) reached 26,649.36 ha year$^{-1}$ for this area. Nonetheless, it is important to note that, in the Caribbean region, a significant percentage of its territory has no forests. Most of the localities of the departments of Cesar, Bolívar, and Sucre have areas of significant deforestation due to the cultivation of palm oil crops and to mining activities. Another source of pressure on the conservation of native forest in Colombia is logging; national timber consumption is estimated to be 4 million m$^3$ year$^{-1}$. Eighty percent of this number comes from natural forests, since the production of reforestation companies meets only 15%–20% of the demand for lumber (Silva, 2006). These operations result in the loss of 128,000 ha year$^{-1}$ of native forest. Illegal crops also have a massive impact on the natural vegetation. Perhaps the most precise estimates come from the localities situated in Middle Magdalena, where native forest losses are estimated at 3,000 ha year$^{-1}$ from this illicit activity. During the period 2002 to 2012, the country's Gross Domestic Product showed exaggerated growth, mainly because of the arrival of investors interested in exploiting oil and mineral resources. Given the increase in geological/mining filings during that time, it is logical to think that none of the sources of pressure mentioned above have decreased but that, on the contrary, new, more-aggressive techniques of deforestation have been added to the traditional sources.

## 5. Conclusion

The utilization rate of forest resources depends essentially on demand. Although in all natural regions, with the exception of the Andean Cordilleras, more than 50% of the natural surface vegetation remains, the weighted value per natural region is approximately 60,000 ha of natural forest that disappears every year because of agriculture and livestock operations or exploitation of mining and energy resources, or 300,000 ha for all regions.

National consumption of timber continues to result in the loss of 128,000 ha of native forest, and around 30,000 ha are destroyed by illegal crops. All of these activities generated a total of 458,000 ha year$^{-1}$ of forest lost or transformed into secondary vegetation. Because reforestation programs are insignificant (as of 2004 they didn't exceed 100,000 ha), if deforestation continues at the current rate, we will deplete our natural capital (forests, biodiversity) in approximately 144 years.

Although there are current data on global deforestation processes based on high-resolution spatial Landsat sensors (Hansen et al., 2013; Kim et al., 2015), they are not calibrated; that is, information on the agents, the transforming processes and their dynamics, and natural vegetation patterns established over a given, varied geography is not known. It is important to emphasize that in spite of this weakness, these data are widely used, for example, in the establishment of baselines for REDD+ projects. The results of these recent studies (Hansen et al., 2013; Kim et al., 2015) should be approached cautiously, because they cannot replace detailed analysis of the coverages and deforestation, based on plots and field observations, which takes into account various social, economic, and ecological contexts.

In Colombia, the central government requires accurate numbers and indicators for its development plans, and thus it is urgent that credibility be established for data on imperative issues such the loss and degradation of vegetation coverage (deforestation). It is essential to have reliable baseline numbers for all statistics for the area covered by forests and different vegetation types in Colombia. This can be achieved with a well-planned forest inventory, with the participation of experts knowledgeable about the issues and the region. Unfortunately, despite the fact that Colombia has the highest biodiversity in the world based on richness of vegetation types, currently we do not have a vegetation map to serve as an appropriate reference for these types of measurements. With political support, these basic tools could be acquired in a short time. With a baseline of real statistical numbers, modern technologies (e.g., satellite images) can make it possible to control processes causing changes in vegetation coverage. As we strive for this essential equilibrium point for reliable statistics, the next step will be to urgently promote the declaration of reserve areas, whether at national, departmental, or municipal levels; increase the extension of various protected areas; encourage growth of the network of private protected reserves; and adopt a strategy to control expansion of the agricultural frontier using land-use management plans and outlines (POT and EOT in Spanish). Another important action is the implementation of programs similar to REDD+, which address economic global realities by promoting compensation for environmental services.

In conclusion, deforestation and degradation are the principal threats to forests and biodiversity in tropical countries. These processes are the largest contributors to climate change in developing countries such as Colombia. Contrary to the conclusions expressed by Sánchez et al. (2012)—that an unexpected trend toward the recovery of forests offers an opportunity to enlarge currently protected areas and promote habitat connectivity—the results of this study show a continuous confusion in the numbers, principally as regards a diminution of the phenomenon (FAO, 2010; IDEAM, 2014). Those flawed results can negatively affect the regions involved, since an apparent disappearance or reduction of the deforestation problem gives the false impression that governmental policies directed toward reducing adverse consequences of climate change are on the right track. This impression must be corrected as soon as possible, or the repercussions will be disastrous for the conserved and vulnerable ecosystems in Colombia, and around the world.

**Acknowledgments**

Part of this work was sponsored by the Bicentennial Project of the National University of Colombia. We thank Andres Avella and Adela Vasquez Lozano for their active participation in various phases of the project. We also want to thank the valuable comments by the Gerardo Aymard, Gustavo Romero and Susan Donoghue as well to the collaboration of Laura Pérez, Vanessa Pérez, Victor Chavez and Henry Arellano-C.

**Author declarations**

The authors make the following declarations about their contributions: the study idea, research design, sampling, cartographic and modeling design, development of CCA and fuzzy logic methodology, data analysis and preparation of the figures of the main article, and supplementary information were provided by HAP. ORC was responsible for the collection and preparation of structural data for conducting the local phytosociological analysis for this study and for providing logistic and field support for the research. HAP wrote the paper and interpreted the results, and both authors wrote the conclusions in global and local contexts and reviewed the manuscript. There is no commercial affiliation between the authors and the entities referred to herein; therefore, no commercial use of the information presented in this document without permission of the principal author is permitted. Creative Commons Attribution-NonCommercial-ShareAlike 4.0 International (CC BY-NC-SA 4.0).

## Appendix A. Mathematical background

*CCA iterative process (Jongman et al., 1995; ter Braak, 1986)*

As a first step, arbitrary values were assigned to localities or plots, and using those values weights of the species ($u_k$) were calculated for the sites based on the point product between the typified biomass values by species and the values for the localities (Equation A1). Then the positions of the sites were recalculated on the basis of the point product of the positions calculated in the previous step and the weights of the species ($x_i$). It is worth mentioning that $y_{ki}$ represents records from the main matrix with typified biomass of species $k$ in plot (relevé) $i$ (Equation A2).

$$u_k = \frac{\sum_{i=1}^{n} y_{ki} x_i}{\sum_{i=1}^{n} y_{ki}} \qquad x_i = \frac{\sum_{k=1}^{m} y_{ki} u_k}{\sum_{k=1}^{m} y_{ki}} \tag{A1,2}$$

These results were used to obtain multiple regression coefficients of the weights of records for the sites with values for typified environmental variables. These data were then used to calculate new records for the sites. According to ter Braak (1986), this is expressed as Equations A3 and A4:

$$b = \left( (z)^t R Z \right)^{-1} \left( (Z)^t R x_i \right) \tag{A3}$$

$$x_i = b_0 + \sum_{j=1}^{q} b_j z_{ij} \quad \text{or in other terms} \quad x_i = zb \quad x_i = b_0 + b_1 z_{1i} + b_2 z_{2i} + b_{...} + b_q z_{qi} \tag{A4}$$

where $z_{ji}$ is the value of environmental variable $j$ at site $i$, $b_j$ is the regression coefficient, and $x_i$ is the value of the composed environmental variable incorporated as a new record for the sites.

The new records had to be typified again (mean of 0, variance of 1) and the iterative process stopped when differences between two final iteration processes were minimal and the correlation maximal; otherwise, the process was repeated from the calculation of Equation A1 ($u_k$). Additional axes were calculated from the regression with the addition of an alternate route, taking uncorrelated values out of the process and eliminating the arc effect caused by the quadratic dependence of two consecutive axes, using axis orthogonalization (not be confused with the orthogonalization of records, explained below).

*Axis orthogonalization (Jongman et al., 1995)*

Additional axes were obtained by linear regression, in which records of sites for the previous axis were designated *f* and records of sites used for this new axis were designated *x* for calculating vector *v* (Equation A5).

$$v = \frac{\sum_{i=1}^{n} y_{si} x_i f_i}{y_{ss}} \quad \text{where} \quad y_{si} = \sum_{k=1}^{m} y_{ki} \quad y_{ss} = \sum_{i=1}^{n} y_{si} \tag{A5}$$

These results were used to calculate a new position of *x* for each species *i* using Equation A6. For remaining axes, the procedure was repeated and sites were typified again.

$$x_{i,new} = x_{i,old} - v f_i \tag{A6}$$

*Standarization (Jongman et al., 1995)*

The standardization process began with the calculation of the center of gravity for **each site (*z*)** (Equation A7):

$$z = \frac{\sum_{i=1}^{n} y_{si} x_i}{y_{ss}} \tag{A7}$$

Then the dispersion of records for the plots was calculated ($S^2$) using Equation A8.

$$s^2 = \frac{\sum_{i=1}^{n} y_{si}(x_i - z)^2}{y_{ss}} \tag{A8}$$

The typified value was calculated using Equation A9.

$$x_{i,new} = \frac{x_{i,old} - z}{s} \tag{A9}$$

The value of *S* is equal to the eigenvalue of each axis when iterations reach convergence.

In the visual implementation of CCA, it was taken into account that values of plots were extracted from species records using the weighted averages algorithm (WA).

*Orthogonalization and calibration results (Peña, 2002)*

The orthogonalization process vectors follow the matrix Equation A10.

$$v = \frac{u.a}{\|a\|^2} a = (uc)c \quad \text{where} \quad c = \frac{a}{\|a\|} \tag{A10}$$

and $\|a\|$ represents the norm or length of $a$ or $\sqrt{a.a}$

*Coverages model validation (Cohen, 1960)*

$$k = \frac{\left(\sum diagX\right) - \left(\sum \frac{\left(\sum_{k=1}^{m} x_{ks}\right)\left(\sum_{i=1}^{n} x_{si}\right)}{N}\right)}{(N) - \left(\sum \frac{\left(\sum_{k=1}^{m} x_{ks}\right)\left(\sum_{i=1}^{n} x_{si}\right)}{N}\right)} \tag{A11}$$

where $X$ is the matrix of confusion.

Preprint: CCA–Fuzzy Land Cover: a new method for classifying vegetation types and coverages and its implications for deforestation analysis    H. Arellano-P. and   J. O. Rangel-Ch. (CC BY-NC-SA 4.0)

Table 1. Physico-environmental variables for vegetation types in pluvial superhumid to semihumid climates in the study region, **and standard deviations for their species.**

| Map symbol | Pattern of vegetation | Altitude ($\sigma$) | $c_1$ | $c_2$ | Incorporation in fuzzy model | Precipitation ($\sigma$) | $c_1$ | $c_2$ | Incorporation in fuzzy model |
|---|---|---|---|---|---|---|---|---|---|
| BHtf2/Jco-Pmu | Possible variant, Jacarando copaiae–Pouterietum multiflorae | 235.94 | 157.10 | 865.30 | Excluded | 230.49 | 2034.80 | 2441.69 | Excluded |
| BHtf2/Jco-Pmu | Jacarando copaiae–Pouterietum multiflorae | 144.10 | 157.10 | 331.38 | Included | 234.55 | 2034.80 | 2441.69 | Included |
| BHtf2/Jco-Pmu/tr | Possible transitional variant, Jacarando copaiae–Pouterietum multiflorae | 308.98 | 819.17 | 865.30 | Excluded | 235.53 | 2177.53 | 2228.65 | Excluded |
| BHmhtf2/Mla-Pma | Marilo laxiflorae–Pentaclethretum macrolobae | 127.77 | 158.44 | 202.98 | | 240.50 | 2135.04 | 2346.12 | |
| BHmhtf1/Par-Vel | Protio aracouchini–Viroletum elongatae | 154.04 | 117.61 | 184.36 | | 229.04 | 1763.49 | 2403.09 | |
| BHtf1/Ean-Cca | Eschweilero antioquensis–Copaiferetum camibar | 218.73 | 684.91 | 729.32 | | 98.15 | 2332.17 | 2434.68 | |
| BHtf1/Twe-Qhu | Tovomito weddellianae–Quercetum humboldtii | 243.20 | 1057.74 | 1088.97 | | 106.55 | 2277.50 | 2314.39 | |
| BHin1/Pde-Tpo | Prestoeo decurrentis–Trichillietum poeppigi | 40.75 | 131.58 | 182.59 | | 357.00 | 2305.28 | 2976.08 | |
| BHri1/Mis-Ppu | Macrolobio ischnocalycis–Peltogynetum purpurea | 106.66 | 159.88 | 310.99 | | 341.80 | 2766.66 | 2928.57 | |
| Bmhtf2/Cpy-Pdo | Cariniano pyriformis–Pentaplarietum doroteae | 166.64 | 95.64 | 134.76 | Included | 288.37 | 1651.28 | 2085.47 | Included |
| Bmhtf2/Mgr-Ama | Mayno grandifoliae–Astrocaryetum malybo | 109.82 | 85.05 | 109.07 | | 179.94 | 1596.13 | 1719.22 | |
| Bhtf2/Cpr-Cpa | Cordietum proctato–panamensis | 37.18 | 81.47 | 131.45 | | 117.46 | 1420.60 | 1517.23 | |
| Bhtf2/Cod-Cpl | Cappari odoratissimatis–Cavanillesietum platanifoliae | 30.96 | 81.11 | 85.79 | | 104.37 | 1459.91 | 1495.11 | |
| Bhtf1/Thi-Spa | Trichilio hirtae–Schizolobietum parahibi | 33.48 | 82.83 | 113.93 | | 204.09 | 1393.32 | 1636.67 | |
| Bhtf2/Tac-Hcr | Trichillio acuminatae–Huretum crepitantis | 63.32 | 5.27 | 131.74 | | 172.21 | 1345.60 | 1535.06 | |
| Bttf3/Asp-Gul | *Acalypha* sp. and *Guazuma ulmifolia* | 53.61 | 14.16 | 142.63 | | 90.36 | 1302.50 | 1476.55 | |
| Map symbol | Pattern of vegetation | Slope ($\sigma$) | $c_1$ | $c_2$ | Incorporation in fuzzy model | Direction ($\sigma$) | $c_1$ | $c_2$ | Incorporation in fuzzy model |
| BHtf2/Jco-Pmu | Possible variant, Jacarando copaiae–Pouterietum multiflorae | 4.56 | 6.79 | 19.92 | | 68.34 | 3.31 | 225.79 | |
| BHtf2/Jco-Pmu | Jacarando copaiae–Pouterietum multiflorae | 4.58 | 6.79 | 19.92 | | 69.93 | 15.47 | 225.79 | Excluded |
| BHtf2/Jco-Pmu/tr | Possible transitional variant, Jacarando copaiae–Pouterietum multiflorae | 4.18 | 15.70 | 17.57 | Excluded | 26.06 | 3.31 | 16.03 | |
| BHmhtf2/Mla-Pma | Marilo laxiflorae–Pentaclethretum macrolobae | 4.24 | 5.75 | 14.34 | | 40.55 | 22.07 | 144.38 | |
| BHmhtf1/Par-Vel | Protio aracouchini–Viroletum elongatae | 4.92 | 7.47 | 20.85 | | 51.67 | 104.12 | 226.18 | Included |
| BHtf1/Ean-Cca | Eschweilero antioquensis–Copaiferetum camibar | 2.60 | 0.71 | 4.33 | | 30.28 | 0.00 | 54.44 | |
| BHtf1/Twe-Qhu | Tovomito weddellianae–Quercetum humboldtii | 2.41 | 12.75 | 14.00 | | 25.89 | 156.67 | 170.23 | |
| BHin1/Pde-Tpo | Prestoeo decurrentis–Trichillietum poeppigi | 2.89 | 0.00 | 7.59 | Included | 33.21 | 0.00 | 103.73 | |
| BHri1/Mis-Ppu | Macrolobio ischnocalycis–Peltogynetum purpurea | 3.27 | 0.00 | 3.56 | | 45.16 | 0.00 | 167.49 | Included |
| Bmhtf2/Cpy-Pdo | Cariniano pyriformis–Pentaplarietum doroteae | 3.81 | 5.56 | 7.67 | Excluded | 30.71 | 70.75 | 122.43 | |
| Bmhtf2/Mgr-Ama | Mayno grandifoliae–Astrocaryetum malybo | 3.47 | 5.22 | 7.16 | | 31.06 | 57.93 | 64.75 | Excluded |
| Bhtf2/Cpr-Cpa | Cordietum proctato–panamensis | 2.06 | 1.92 | 5.45 | Included | 33.90 | 44.37 | 164.81 | Included |
| Bhtf2/Cod-Cpl | Cappari odoratissimatis–Cavanillesietum platanifoliae | 1.91 | 3.78 | 5.16 | Excluded | 45.86 | 66.16 | 108.84 | |
| Bhtf1/Thi-Spa | Trichilio hirtae–Schizolobietum parahibi | 3.25 | 3.87 | 10.71 | Included | 59.98 | 29.87 | 262.63 | Excluded |
| Bhtf2/Tac-Hcr | Trichillio acuminatae–Huretum crepitantis | 4.83 | 11.62 | 15.38 | Excluded | 41.79 | 95.18 | 188.68 | Included |
| Bttf3/Asp-Gul | *Acalypha* sp. and *Guazuma ulmifolia* | 3.76 | 0.72 | 15.59 | Included | 57.10 | 0.00 | 268.12 | Excluded |

| Map symbol | Pattern of vegetation | TCI (σ) | c1 | c2 | Incorporation in fuzzy model |
|---|---|---|---|---|---|
| BHtf2/Jco-Pmu | Possible variant, Jacarando copaiae–Pouterietum multiflorae | 1.21 | 2.93 | 5.08 | |
| BHtf2/Jco-Pmu | Jacarando copaiae–Pouterietum multiflorae | 1.09 | 2.93 | 4.30 | |
| BHtf2/Jco-Pmu/tr | Possible transitional variant, Jacarando copaiae–Pouterietum multiflorae | 1.31 | 4.59 | 5.08 | Excluded |
| BHmhtf2/Mla-Pma | Marilo laxiflorae–Pentaclethretum macrolobae | 1.43 | 2.12 | 5.40 | |
| BHmhtf1/Par-Vel | Protio aracouchini–Viroletum elongatae | 1.42 | 3.81 | 7.23 | |
| BHtf1/Ean-Cca | Eschweilero antioquensis–Copaiferetum camibar | 2.05 | 2.69 | 10.52 | Included |
| BHtf1/Twe-Qhu | Tovomito weddellianae–Quercetum humboldtii | 0.93 | 4.92 | 6.01 | Excluded |
| BHin1/Pde-Tpo | Prestoeo decurrentis–Trichillietum poeppigi | 2.01 | 6.91 | 13.73 | Included |
| BHri1/Mis-Ppu | Macrolobio ischnocalycis–Peltogynetum purpurea | 1.69 | 3.04 | 8.29 | |
| Bmhtf2/Cpy-Pdo | Cariniano pyriformis–Pentaplarietum doroteae | 1.29 | 3.94 | 5.53 | Excluded |
| Bmhtf2/Mgr-Ama | Mayno grandifoliae–Astrocaryetum malybo | 0.97 | 2.71 | 4.90 | |
| Bhtf2/Cpr-Cpa | Cordietum proctato–panamensis | 1.10 | 3.47 | 8.01 | Included |
| Bhtf2/Cod-Cpl | Cappari odoratissimatis–Cavanillesietum platanifoliae | 1.57 | 5.66 | 6.61 | |
| Bhtf1/Thi-Spa | Trichilio hirtae–Schizolobietum parahibi | 1.02 | 3.94 | 5.90 | Excluded |
| Bhtf2/Tac-Hcr | Trichillio acuminatae–Huretum crepitantis | 1.18 | 5.04 | 7.50 | |
| Bttf3/Asp-Gul | *Acalypha* sp. and *Guazuma ulmifolia* | 1.11 | 3.26 | 7.26 | Included |

Table 2. Physico-environmental variables for vegetation types in dry climates in the study region, **and standard deviations for their species.**

| Map symbol | Pattern of vegetation | Altitude (σ) | c1 | c2 | Incorporation in fuzzy model | Precipitation (σ) | c1 | c2 | Incorporation in fuzzy model |
|---|---|---|---|---|---|---|---|---|---|
| Bhtf2/Sma-Cpl | Sabali mauritiiformis–Cavanillesietum platanifoliae* | 25.31378 | 39.01875 | 104.61859 | | 140.33765 | 1354.38137 | 1389.03046 | |
| Bhttf2/Ctr-Aas | Cinnamomo triplinervis–Apeibetum asperae | 29.45018 | 32.18314 | 83.16916 | | 251.93008 | 1069.52031 | 2094.59435 | |
| Bttf2/Cvi-Abu | Cochlospermo vitifoli–Attaleetum butyraceae | 32.60334 | 35.23321 | 51.45752 | | 143.45696 | 1198.84014 | 1268.35283 | |
| Bmhtf2/Oma-Cpy | Oenocarpus mapora y Cariniana pyriformis | 3.15989 | 49.82548 | 49.82548 | | 68.47183 | 1881.18488 | 1881.18488 | |
| Bmhtf2/Vca-Aac | Viticis capitatae–Acrocomietum aculeatae | 18.19858 | 21.28044 | 57.34496 | | 239.35544 | 1838.90362 | 2342.75950 | |
| Bmhtf2/Xar-Tgu | Xylopio aromaticae–Tapiriretum guianensis | 10.28761 | 15.72607 | 37.10931 | Included | 394.70845 | 2596.17261 | 2778.57854 | Included |
| Bttf2/Apu-Pla | Annono punicifoliae–Pithecellobietum lanceolati | 34.57615 | 1.94081 | 20.99449 | | 324.95623 | 1318.25906 | 1570.80687 | |
| Bmhtf2/Spa-Tro | Symmerio paniculatae–Tabebuietum roseae | 8.06323 | 11.13861 | 22.59927 | | 271.74251 | 1472.70910 | 2047.82426 | |
| Bttf2/Call-Abu | Cordio alliodorae–Attaleetum butyraceae | 75.64094 | 0.44673 | 213.12866 | | 78.34502 | 1034.15603 | 1167.80043 | |
| Bttf2/Ain-Agr | Adenocalymno inundati–Astronietum graveolentis | 7.59708 | 19.89434 | 40.55149 | | 161.84819 | 1191.41445 | 1280.72983 | |
| Bmhtf2/Cvi-Abu | Cochlospermo vitifoli–Mataybetum camptoneurae | 16.41811 | 20.53430 | 40.73143 | | 61.57849 | 1904.15057 | 2159.75485 | |
| Bmhin2/Cco-Ahu | Coccolobo costatae–Acacietum huilanae | 2.76948 | 13.08916 | 21.89450 | | 154.65301 | 2065.66284 | 2075.43686 | |
| Btin2/Ctr-Ssa | Caseario tremulae–Samaneetum samanis | 8.60484 | 6.51304 | 7.29634 | | 516.42871 | 1356.64033 | 1381.04986 | |
| Btin2/Mar | *Montrichardia arborescens* community | 0.10000 | 0.00000 | 50.00000 | | | | | |

| Map symbol | Pattern of vegetation | Direction (σ) | c1 | c2 | Incorporation in fuzzy model | TCI (σ) | c1 | c2 | Incorporation in fuzzy model |
|---|---|---|---|---|---|---|---|---|---|
| Bhtf2/Sma-Cpl | Sabali mauritiiformis–Cavanillesietum platanifoliae | 87.80497705 | 0 | 188.8708257 | | 4.216926578 | 4.659214025 | 18.08620112 | Included |
| Bhttf2/Ctr-Aas | Cinnamomo triplinervis–Apeibetum asperae | 49.66476426 | 0 | 145.2842469 | Included | 1.352684825 | 4.354538258 | 5.670216673 | Excluded |
| Bttf2/Cvi-Abu | Cochlospermo vitifoli–Attaleetum butyraceae | 88.72860381 | 75.24230907 | 198.6299808 | | 2.057999666 | 5.096231386 | 7.254689188 | |
| Bmhtf2/Oma-Cpy | *Oenocarpus mapora* and *Cariniana pyriformis* | 21.40594864 | 0 | 0 | Excluded | 0.504310949 | 7.043783137 | 7.043783137 | |
| Bmhtf2/Vca-Aac | Viticis capitatae–Acrocomietum aculeatae | 46.40748347 | 0 | 133.2691668 | Included | 2.019222744 | 3.581700528 | 8.429050201 | Included |
| Bmhtf2/Xar-Tgu | Xylopio aromaticae–Tapiriretum guianensis | 44.66268452 | 4.529738943 | 45.44794723 | | 1.268333548 | 3.767990969 | 5.127140782 | |
| Bttf2/Apu-Pla | Annono punicifoliae–Pithecellobietum lanceolati | 56.22392638 | 0 | 155.2388487 | | 1.2190642 | 2.371510924 | 4.974075101 | |
| Bmhtf2/Spa-Tro | Symmerio paniculatae–Tabebuietum roseae | 40.20035398 | 0 | 41.71400602 | Included | 1.266863778 | 3.331689934 | 5.281047818 | |
| Bttf2/Call-Abu | Cordio alliodorae–Attaleetum butyraceae | 24.56141734 | 0 | 70.33163907 | | 1.179821565 | 1.583222018 | 3.192748814 | Excluded |
| Bttf2/Ain-Agr | Adenocalymno inundati–Astronietum graveolentis | 77.3667419 | 0 | 345.5145942 | | 0.877645485 | 3.158879385 | 5.987375506 | |
| Bmhtf2/Cvi-Abu | Cochlospermo vitifoli–Mataybetum camptoneurae | 101.0665036 | 0 | 259.2546849 | | 0.830529182 | 1.575607389 | 4.34522097 | |
| Bmhin2/Cco-Ahu | Coccolobo costatae–Acacietum huilanae | 81.11679202 | 33.48201377 | 242.1972637 | Excluded | 0.947940351 | 6.164561006 | 8.468127189 | |
| Btin2/Ctr-Ssa | Caseario tremulae–Samaneetum samanis | 37.63032025 | 26.03764694 | 32.27110109 | | 0.884049774 | 5.0900064 | 5.118788206 | |
| Btin2/Mar | *Montrichardia arborescens* community | 0 | 0 | 0 | | 4.216926578 | 4.659214025 | 18.08620112 | Included |
| Map symbol | Pattern of vegetation | Slope (σ) | $c_1$ | $c_2$ | Incorporation in fuzzy model | | | | |
| Btin2/Mar | *Montrichardia arborescens* community | 0.1 | 0 | 3.563000658 | Included | | | | |

* The association Sabali mauritiiformis–Cavanillesietum platanifoliae can also be found in semihumid climates.

Table 3. Statistical results for physico-environmental variables used in the ordination of species and plots by CCA in the study region.

| Combination of environmental variables explored | General vegetation pattern | Data Correction | Total variance | Eigenvalues | | | MONTECARLO for Eigenvalues (Average, Min, Max, p) | | | | Variable correlated (Axis 1, 2, 3) | Percentage explained by Axes | | | Cumulative percentage | | | Pearson Correlation between species and environment variables | | | MONTECARLO for correlation (Average, Min, Max, p) | | | |
|---|---|---|---|---|---|---|---|---|---|---|---|---|---|---|---|---|---|---|---|---|---|---|---|---|
| TCI-Direction-Slope | Tropical forests in humid climates from superhumid-pluvial to semihumid climates. A humid-dry transitional group is included (62 Plots Sp-703). | Original, same as the publication or source data. | 31.3781 | 0.688 | 0.610 | 0.514 | 0.572 0.482 0.677 0.0010 | 0.510 0.411 0.606 | 0.448 0.320 0.544 | | Slope / TCI / Direction | 2.2 | 1.9 | 1.6 | 2.2 | 4.1 | 5.8 | 0.947 0.910 0.902 | | | 0.941 0.895 0.974 0.3594 | 0.926 0.860 0.974 | 0.905 0.828 0.956 | |
| Altitude-Slope-TCI | | | | 0.904 | 0.662 | 0.592 | 0.575 0.470 0.733 0.0010 | 0.508 0.403 0.620 | 0.442 0.264 0.576 | | Altitude / Slope / TCI | 2.9 | 2.1 | 1.9 | 2.9 | 5.0 | 6.9 | 0.989 0.945 0.897 | | | 0.943 0.890 0.982 0.0010 | 0.926 0.852 0.977 | 0.905 0.781 0.955 | |
| TCI-Direction-Slope | Tropical forests in dry climates. A humid-dry transitional group is included (50 Plots Sp-309). | | 24.654 | 0.679 | 0.405 | 0.306 | 0.660 0.463 0.868 0.3833 | 0.538 0.291 0.745 | 0.410 0.141 0.644 | | Direction / TCI / Slope | 2.8 | 1.6 | 1.2 | 2.8 | 4.4 | 5.6 | 0.907 0.759 0.759 | | | 0.932 0.828 0.986 0.8311 | 0.891 0.694 0.978 | 0.820 0.499 0.956 | |
| Altitude-Slope-TCI | | | | 0.734 | 0.457 | 0.285 | 0.660 0.453 0.879 0.1467 | 0.541 0.303 0.772 | 0.407 0.122 0.692 | | Altitude / TCI / Slope | 3.0 | 1.9 | 1.2 | 3.0 | 4.8 | 6.0 | 0.937 0.776 0.776 | | | 0.934 0.800 0.989 0.4978 | 0.894 0.736 0.973 | 0.822 0.535 0.958 | |
| TCI-Direction-Slope | Tropical forests in humid climates from superhumid-pluvial to semihumid climates. The humid-dry transitional group is excluded (60 Plots Sp-713). | Some location data corrected by approximation and calibration. | 29.7634 | 0.638 | 0.561 | 0.492 | 0.561 0.469 0.660 0.0100 | 0.499 0.408 0.602 | 0.438 0.323 0.532 | | Slope / TCI / Direction | 2.1 | 1.9 | 1.7 | 2.1 | 4.0 | 5.7 | 0.928 0.897 0.875 | | | 0.936 0.886 0.974 0.7457 | 0.920 0.866 0.964 | 0.899 0.818 0.972 | |
| Altitude-Precipitation-TCI | | | | 0.895 | 0.823 | 0.511 | 0.565 0.479 0.683 0.0010 | 0.499 0.377 0.596 | 0.433 0.297 0.541 | | Altitude / Precipitation / TCI | 3.0 | 2.8 | 1.7 | 3.0 | 5.8 | 7.5 | 0.982 0.969 0.890 | | | 0.938 0.890 0.977 0.0010 | 0.921 0.836 0.974 | 0.900 0.772 0.957 | |
| TCI-Direction-Slope | Tropical forests in dry climates. The humid-dry transitional group is excluded (52 Plots Sp-313). | | 28.8046 | 0.660 | 0.484 | 0.421 | 0.639 0.510 0.811 0.3093 | 0.559 0.434 0.695 | 0.478 0.261 0.615 | | Direction / Slope / TCI | 2.3 | 1.7 | 1.5 | 2.3 | 4.0 | 5.4 | 0.938 0.842 0.803 | | | 0.952 0.891 0.986 0.8318 | 0.932 0.852 0.979 | 0.905 0.730 0.962 | |
| Altitude-Slope-TCI | | | | 0.710 | 0.663 | 0.462 | 0.635 0.509 0.804 0.0621 | 0.557 0.431 0.717 | 0.475 0.281 0.618 | | Precipitation / Altitude / TCI | 2.5 | 2.3 | 1.6 | 2.5 | 4.8 | 6.4 | 0.946 0.942 0.843 | | | 0.950 0.885 0.987 0.6476 | 0.932 0.840 0.977 | 0.904 0.744 0.968 | |
| TCI-Direction-Slope | Tropical forests in humid climates from superhumid-pluvial to dry (112 Plots Sp-938). | Some location data corrected by approximation and calibration. | 55.5331 | 0.617 | 0.539 | 0.508 | 0.546 0.457 0.642 0.0130 | 0.498 0.430 0.578 | 0.449 0.362 0.529 | | Slope / TCI / Direction | 1.1 | 1.0 | 0.9 | 1.1 | 2.1 | 3.0 | 0.896 0.885 0.872 | | | 0.928 0.881 0.961 0.9940 | 0.913 0.864 0.948 | 0.896 0.833 0.942 | |

| Combination | Description | Value | Col A | Col B | Col C | V1 | V2 | V3 | p | Variable | N1 | N2 | N3 | N4 | N5 | N6 | M1 | M2 | M3 | R1 | R2 | R3 | p2 |
|---|---|---|---|---|---|---|---|---|---|---|---|---|---|---|---|---|---|---|---|---|---|---|---|
| Altitude-Slope-TCI | | | 0.800 | 0.790 | 0.471 | 0.548 | 0.480 | 0.646 | 0.0010 | Altitude | | | | | | | | | | 0.928 | 0.873 | 0.962 | 0.0120 |
| | | | | | | 0.496 | 0.423 | 0.585 | | Slope | 1.4 | 1.4 | 0.8 | 1.4 | 2.9 | 3.7 | 0.954 | 0.967 | 0.870 | 0.912 | 0.868 | 0.955 | |
| | | | | | | 0.447 | 0.359 | 0.526 | | TCI | | | | | | | | | | 0.896 | 0.820 | 0.943 | |
| Altitude-Precipitation-Slope | | | 0.907 | 0.829 | 0.592 | 0.574 | 0.467 | 0.693 | 0.0010 | Precipitation | | | | | | | | | | 0.943 | 0.891 | 0.986 | 0.0010 |
| | | | | | | 0.511 | 0.391 | 0.601 | | Altitude | 2.9 | 2.6 | 1.9 | 2.9 | 5.5 | 7.4 | 0.987 | 0.977 | 0.922 | 0.926 | 0.858 | 0.969 | |
| | | | | | | 0.447 | 0.306 | 0.545 | | Slope | | | | | | | | | | 0.907 | 0.807 | 0.962 | |
| Altitude-Precipitation-TCI | | | 0.901 | 0.825 | 0.501 | 0.573 | 0.460 | 0.740 | 0.0010 | Altitude | | | | | | | | | | 0.943 | 0.890 | 0.983 | 0.0010 |
| | | | | | | 0.510 | 0.418 | 0.617 | | Precipitation | 2.9 | 2.6 | 1.6 | 2.9 | 5.5 | 7.1 | 0.985 | 0.973 | 0.873 | 0.926 | 0.852 | 0.967 | |
| | | | | | | 0.444 | 0.292 | 0.537 | | TCI | | | | | | | | | | 0.905 | 0.817 | 0.964 | |
| Altitude-Precipitation-Direction | Tropical forests in humid climates from superhumid-pluvial to semihumid climates. A humid-dry transitional group is included (62 Plots Sp-703). | 31.3781 | 0.900 | 0.826 | 0.472 | 0.572 | 0.461 | 0.742 | 0.0010 | Altitude | | | | | | | | | | 0.942 | 0.888 | 0.984 | 0.0010 |
| | | | | | | 0.510 | 0.410 | 0.607 | | Precipitation | 2.9 | 2.6 | 1.5 | 2.9 | 5.5 | 7.0 | 0.985 | 0.972 | 0.869 | 0.926 | 0.851 | 0.968 | |
| | | | | | | 0.448 | 0.341 | 0.537 | | Direction | | | | | | | | | | 0.907 | 0.830 | 0.955 | |
| Slope-Precipitation-TCI | | | 0.893 | 0.631 | 0.492 | 0.570 | 0.455 | 0.677 | 0.0010 | Precipitation | | | | | | | | | | 0.941 | 0.886 | 0.969 | 0.0010 |
| | | | | | | 0.509 | 0.440 | 0.601 | | Slope | 2.8 | 2.0 | 1.6 | 2.8 | 4.9 | 6.4 | 0.982 | 0.922 | 0.875 | 0.926 | 0.870 | 0.965 | |
| | | | | | | 0.448 | 0.337 | 0.531 | | TCI | | | | | | | | | | 0.906 | 0.839 | 0.961 | |
| Altitude-Slope-TCI | Some location data corrected by approximation and calibration. | | 0.881 | 0.592 | 0.565 | 0.574 | 0.479 | 0.716 | 0.0010 | Altitude | | | | | | | | | | 0.942 | 0.888 | 0.986 | 0.0010 |
| | | | | | | 0.509 | 0.413 | 0.627 | | Slope | 2.8 | 1.9 | 1.8 | 2.8 | 4.7 | 6.5 | 0.988 | 0.918 | 0.886 | 0.926 | 0.867 | 0.972 | |
| | | | | | | 0.445 | 0.298 | 0.545 | | TCI | | | | | | | | | | 0.906 | 0.807 | 0.963 | |
| TCI-Direction-Slope | | | 0.652 | 0.573 | 0.497 | 0.570 | 0.485 | 0.671 | 0.0070 | Slope | | | | | | | | | | 0.941 | 0.892 | 0.976 | 0.7437 |
| | | | | | | 0.510 | 0.425 | 0.593 | | TCI | 2.1 | 1.8 | 1.6 | 2.1 | 3.9 | 5.5 | 0.933 | 0.878 | 0.884 | 0.925 | 0.870 | 0.963 | |
| | | | | | | 0.450 | 0.351 | 0.554 | | Direction | | | | | | | | | | 0.908 | 0.821 | 0.950 | |
| Altitude-Precipitation-TCI | | | 0.737 | 0.678 | 0.581 | 0.649 | 0.485 | 0.857 | 0.0511 | Altitude | | | | | | | | | | 0.954 | 0.892 | 0.985 | 0.5495 |
| | | | | | | 0.568 | 0.434 | 0.731 | | Precipitation | 2.6 | 2.4 | 2.1 | 2.6 | 5.0 | 7.1 | 0.954 | 0.930 | 0.899 | 0.935 | 0.863 | 0.980 | |
| | | | | | | 0.480 | 0.285 | 0.621 | | TCI | | | | | | | | | | 0.903 | 0.732 | 0.974 | |
| Altitude-Precipitation-Direction | Tropical forests in dry climates. A humid-dry transitional group is included (50 Plots Sp-309). | 28.2498 | 0.720 | 0.688 | 0.606 | 0.649 | 0.533 | 0.807 | 0.0561 | Altitude | | | | | | | | | | 0.954 | 0.892 | 0.984 | 0.5516 |
| | | | | | | 0.570 | 0.435 | 0.700 | | Precipitation | 2.5 | 2.4 | 2.1 | 2.5 | 5.0 | 7.1 | 0.953 | 0.937 | 0.924 | 0.934 | 0.872 | 0.973 | |
| | | | | | | 0.488 | 0.290 | 0.613 | | Direction | | | | | | | | | | 0.907 | 0.743 | 0.969 | |
| Altitude-Slope-TCI | | | 0.743 | 0.587 | 0.405 | 0.657 | 0.466 | 0.886 | 0.0761 | Altitude | | | | | | | | | | 0.956 | 0.857 | 0.989 | 0.7017 |
| | | | | | | 0.568 | 0.391 | 0.765 | | TCI | 2.6 | 2.1 | 1.4 | 2.6 | 4.7 | 6.1 | 0.950 | 0.905 | 0.788 | 0.936 | 0.786 | 0.983 | |
| | | | | | | 0.473 | 0.224 | 0.645 | | Slope | | | | | | | | | | 0.902 | 0.712 | 0.977 | |
| Altitude-Precipitation- | | | 0.728 | 0.681 | 0.430 | 0.654 | 0.485 | 0.881 | 0.1041 | Altitude | 2.6 | 2.4 | 1.5 | 2.6 | 5.0 | 6.5 | 0.954 | 0.932 | 0.803 | 0.955 | 0.889 | 0.988 | 0.5536 |

| | | | | | | | | | | | | | | | | | | | | |
|---|---|---|---|---|---|---|---|---|---|---|---|---|---|---|---|---|---|---|---|---|
| Slope | | | | 0.570 | 0.380 | 0.752 | | Precipitation | | | | | | | | | | 0.936 | 0.777 | 0.986 |
| | | | | 0.478 | 0.294 | 0.641 | | Slope | | | | | | | | | | 0.903 | 0.741 | 0.977 |
| | | | | 0.648 | 0.477 | 0.820 | 0.2242 | Precipitation | | | | | | | | | | 0.953 | 0.881 | 0.989 | 0.8899 |
| Slope-Precipitation-TCI | 0.684 | 0.625 | 0.407 | 0.568 | 0.431 | 0.730 | | TCI | 2.4 | 2.2 | 1.4 | 2.4 | 4.6 | 6.1 | 0.934 | 0.903 | 0.791 | 0.934 | 0.870 | 0.982 |
| | | | | 0.484 | 0.325 | 0.628 | | Slope | | | | | | | | | | 0.904 | 0.770 | 0.968 |
| | | | | 0.649 | 0.506 | 0.826 | 0.5015 | TCI | | | | | | | | | | 0.953 | 0.890 | 0.992 | 0.9930 |
| TCI-Direction-Slope | 0.646 | 0.614 | 0.396 | 0.564 | 0.420 | 0.711 | | Direction | 2.3 | 2.2 | 1.4 | 2.3 | 4.5 | 5.9 | 0.911 | 0.919 | 0.786 | 0.933 | 0.828 | 0.976 |
| | | | | 0.478 | 0.303 | 0.621 | | Slope | | | | | | | | | | 0.903 | 0.776 | 0.968 |

Table 4. Consolidated gross values.

| Year 1987, general coverage | Year 2010, general coverage | Phenomenon related to change | Area in ha | Percentage | Change in ha year$^{-1}$ (1987–2010) | Change in ha year$^{-1}$ (2000–2010), Hansen et al. (2013) | Change in ha year$^{-1}$ (2010–2013), Hansen et al. (2013) |
|---|---|---|---|---|---|---|---|
| Grasslands | Grasslands | Constant | 214398.68 | 6.60 | | | |
| Forest | Forest | | 1075587.52 | 33.12 | | | |
| Shrubs | Shrubs | | 240720.44 | 7.41 | | | |
| Deforestation areas | Deforestation areas | | 668885.52 | 20.60 | | | |
| Subtotal for constant coverages in the period evaluated | | | 2199592.16 | 67.74 | | | |
| Deforestation **Buffer** | Grasslands | Grasslands buffer | 466.08 | 0.01 | | | |
| Grasslands **Buffer** | Deforestation | | 312.24 | 0.01 | | | |
| Total for constant and relatively constant coverages during the period evaluated | | | 2200370.48 | 67.76 | | | |
| Forest | Deforestation | Gross deforestation | 147384.64 | 4.54 | | | |
| Shrubs | | | 111709.84 | 3.44 | | | |
| Subtotal for the phenomenon of gross deforestation itself | | | 259094.48 | 7.98 | 11264.98 | | |
| Forest | Grasslands | Gross deforestation 2 | 99994.20 | 3.08 | | | |
| Shrubs | | | 63122.52 | 1.94 | | | |
| Subtotal for the phenomenon of gross deforestation type two | | | 163116.72 | 5.02 | 7092.03 | | |
| Total for the phenomenon of gross deforestation | | | 422211.20 | 13.00 | 18357.01 | | |
| Forest | Shrubs | Gross degradation | 189945.84 | 5.85 | 8258.51 | | |
| Total for the phenomenon of transformation (gross loss) | | | 612157.04 | 18.85 | 26615.52 | 9405.13 | 1671.41 |
| Shrubs | Secondary forest, poor structure | | 198137.52 | 6.10 | | | |
| Grasslands | | | 23797.84 | 0.73 | | | |
| Grasslands | Shrubs | Gross recuperation | 46980.04 | 1.45 | | | |
| Deforestation | | | 110765.76 | 3.41 | | | |
| Deforestation | Secondary forest, very poor structure | | 54984.44 | 1.69 | | | |
| Total for the recovery process with poor vegetation structure (gross gain) | | | 434665.60 | 13.39 | 18898.50 | | |

Table 5. Net gains and losses for certain vegetation types in the study area.

| Map symbol | Vegetation pattern | Number symbol | Association | 1987 Area (ha) | 2010 Area (ha) | Net loss (1987–2010) | Net gain (1987–2010) |
|---|---|---|---|---|---|---|---|
| BHtf2/Jco-Pmu | | 1 | Jacarando copaiae–Pouterietum multiflorae | 216552.12 | 215242.52 | 1309.6 | -- |
| BHmhtf2/Mla-Pma | | 2 | Marilo laxiflorae–Pentaclethretum macrolobae | 7843.44 | 8466.4 | -- | 622.96 |
| BHmhtf1/Par-Vel | | 3 | Protio aracouchini–Viroletum elongatae | 39757.24 | 35246.2 | 4511.04 | -- |
| BHtf1/Ean-Cca | Primary forest | 4 | Eschweilero antioquensis–Copaiferetum camibar | 90395.16 | 90421.16 | -- | 26 |
| BHtf1/Twe-Qhu | | 5 | Tovomito weddellianae–Quercetum humboldtii | 35652.36 | 36087.28 | -- | 434.92 |
| BHin1/Pde-Tpo | | 6 | Prestoeo decurrentis–Trichillietum poeppigi | 78109 | 76987.72 | 1121.28 | -- |
| BHri1/Mis-Ppu | | 7 | Macrolobio ischnocalycis-Peltogynetum purpurea | 70863.64 | 71897.96 | -- | 1034.32 |
| Total of change for vegetation types dominated by primary forests | | | | | | 6941.92 | 2118.2 |
| Bmhtf2/Cpy-Pdo | | 8 | Cariniano pyriformis–Pentaplarietum doroteae | 9061.12 | 6681.04 | 2380.08 | -- |
| Bmhtf2/Mgr-Ama | | 9 | Mayno grandifoliae–Astrocaryetum malybo | 11626.04 | 8272.2 | 3353.84 | -- |
| Bhtf2/Cpr-Cpa | | 10 | Cordietum proctato–panamensis | 1148.04 | 764.32 | 383.72 | -- |
| Bhtf2/Cod-Cpl | | 11 | Cappari odoratissimatis–Cavanillesietum platanifoliae | 170.68 | 134.96 | 35.72 | -- |
| Bhtf1/Thi-Spa | | 12 | Trichilio hirtae–Schizolobietum parahibi | 10081.64 | 7998.16 | 2083.48 | -- |
| Bhtf2/Tac-Hcr | | 13 | Trichillio acuminatae–Huretum crepitantis | 36278.04 | 30635.8 | 5642.24 | -- |
| Bttf3/Asp-Gul | | 14 | *Acalypha* sp.and *Guazuma ulmifolia* | 43867.92 | 33599.72 | 10268.2 | -- |
| Bhtf2/Sma-Cpl | | 15 | Sabali mauritiiformis–Cavanillesietum platanifoliae | 35913.92 | 23021.4 | 12892.52 | -- |
| Bhttf2/Ctr-Aas | | 16 | Cinnamomo triplinervis–Apeibetum asperae | 255637.44 | 187514.6 | 68122.84 | -- |
| Bttf2/Cvi-Abu | Forests with different levels of intervention | 17 | Cochlospermo vitifoli–Attaleetum butyraceae | 13072.6 | 12596.12 | 476.48 | -- |
| Bmhtf2/Oma-Cpy | | 18 | *Oenocarpus mapora* and *Cariniana pyriformis* | 7986.24 | 7351.4 | 634.84 | -- |
| Bmhtf2/Vca-Aac | | 19 | Viticis capitatae–Acrocomietum aculeatae | 43204.28 | 37988.64 | 5215.64 | -- |
| Bmhtf2/Xar-Tgu | | 20 | Xylopio aromaticae–Tapiriretum guianensis | 175259.2 | 182073.48 | -- | 6814.28 |
| Bttf2/Apu-Pla | | 21 | Annono puniciofoliae–Pithecellobietum lanceolati | 22531.8 | 17124.84 | 5406.96 | -- |
| Bmhtf2/Spa-Tro | | 22 | Symmerio paniculatae–Tabebuietum roseae | 106722.32 | 99969 | 6753.32 | -- |
| Bttf2/Call-Abu | | 23 | Cordio alliodorae–Attaleetum butyraceae | 115861.88 | 90287.04 | 25574.84 | -- |
| Bttf2/Ain-Agr | | 24 | Adenocalymno inundati–Astronietum graveolentis | 24978.44 | 15841.8 | 9136.64 | -- |
| Bmhtf2/Cvi-Abu | | 25 | Cochlospermo vitifoli–Maytabetum camptoneurae | 46100.88 | 41990 | 4110.88 | -- |
| Bmhin2/Cco-Ahu | | 26 | Coccolobo costatae–Acacietum huilanae | 14059.32 | 14147.84 | -- | 88.52 |
| Btin2/Ctr-Ssa | | 27 | Caseario tremulae–Samaneetum samanis | 177.48 | 165.76 | 11.72 | -- |
| Total of change for vegetation types dominated by forests with different levels of intervention | | | | | | 162483.96 | 6902.8 |
| Total change for all types of vegetation dominated by forests | | | | | | 169425.88 | 9021 |
| MHtf2/Jco-Pmu | | 29 | Jacarando copaiae–Pouterietum multiflorae | 26188.04 | 24555.12 | 1632.92 | -- |
| MHmhtf2/Mla-Pma | | 30 | Marilo laxiflorae–Pentaclethretum macrolobae | 1183.92 | 630 | 553.92 | -- |
| MHmhtf1/Par-Vel | Tall and medium shrubs | 31 | Protio aracouchini–Viroletum elongatae | 9063.04 | 10153.76 | -- | 1090.72 |
| MHtf1/Ean-Cca | | 32 | Eschweilero antioquensis–Copaiferetum camibar | 12.6 | 7.08 | 5.52 | -- |
| MHtf1/Twe-Qhu | | 33 | Tovomito weddellianae–Quercetum humboldtii | 21.4 | 41.4 | -- | 20 |
| MHin1/Pde-Tpo | | 34 | Prestoeo decurrentis–Trichillietum poeppigi | 3443.36 | 4731.04 | -- | 1287.68 |
| MHri1/Mis-Ppu | | 35 | Macrolobio ischnocalycis–Peltogynetum purpurea | 5290.4 | 6313.24 | -- | 1022.84 |

| | | | | | | |
|---|---|---|---|---|---|---|
| Mmhtf2/Cpy-Pdo | 36 | Cariniano pyriformis–Pentaplarietum doroteae | 1534.2 | 2549.84 | -- | 1015.64 |
| Mmhtf2/Mgr-Ama | 37 | Mayno grandifoliae–Astrocaryetum malybo | 5072.92 | 5453.68 | -- | 380.76 |
| Mhtf2/Cpr-Cpa | 38 | Cordietum proctato-panamensis | 1399.64 | 1387.92 | 11.72 | -- |
| Mhtf2/Cod-Cpl | 39 | Cappari odoratissimatis–Cavanillesietum platanifoliae | 250.84 | 225.92 | 24.92 | -- |
| Mhtf1/Thi-Spa | 40 | Trichilio hirtae–Schizolobietum parahibi | 12452.4 | 11679.64 | 772.76 | -- |
| Mhtf2/Tac-Hcr | 41 | Trichillio acuminatae–Huretum crepitantis | 32260.04 | 26895.6 | 5364.44 | -- |
| Mttf3/Asp-Gul | 42 | *Acalypha* sp.and *Guazuma ulmifolia* | 44392.2 | 35262.28 | 9129.92 | -- |
| Mhtf2/Sma-Cpl | 43 | Sabali mauritiiformis–Cavanillesietum platanifoliae | 36735.64 | 32971.32 | 3764.32 | -- |
| Mhttf2/Ctr-Aas | 44 | Cinnamomo triplinervis–Apeibetum asperae | 150862.88 | 173626.16 | -- | 22763.28 |
| Mttf2/Cvi-Abu | 45 | Cochlospermo vitifoli–Attaleetum butyraceae | 12152.12 | 12884.44 | -- | 732.32 |
| Mmhtf2/Oma-Cpy | 46 | *Oenocarpus mapora* and *Cariniana pyriformis* | 865.96 | 1358.88 | -- | 492.92 |
| Mmhtf2/Vca-Aac | 47 | Viticis capitatae–Acrocomietum aculeatae | 25719.68 | 24855.64 | 864.04 | -- |
| Mmhtf2/Xar-Tgu | 48 | Xylopio aromaticae–Tapiriretum guianensis | 63542.68 | 47805.92 | 15736.76 | -- |
| Mttf2/Apu-Pla | 49 | Annono punicifoliae-Pithecellobietum lanceolati | 19960.16 | 17485.84 | 2474.32 | -- |
| Mmhtf2/Spa-Tro | 50 | Symmerio paniculatae–Tabebuietum roseae | 54922.56 | 39228.36 | 15694.2 | -- |
| Mttf2/Call-Abu | 51 | Cordio alliodorae–Attaleetum butyraceae | 63918.52 | 73865.16 | -- | 9946.64 |
| Mttf2/Ain-Agr | 52 | Adenocalymno inundati–Astronietum graveolentis | 15755.8 | 13649.76 | 2106.04 | -- |
| Mmhtf2/Cvi-Abu | 53 | Cochlospermo vitifoli–Maraybetum camptoneurae | 19927.88 | 16062.12 | 3865.76 | -- |
| Mmhin2/Cco-Ahu | 54 | Coccolobo costatae–Acacietum huilanae | 6668.48 | 4632.04 | 2036.44 | -- |
| Mtin2/Ctr-Ssa | 55 | Caseario tremulae–Samaneetum samanis | 92.96 | 99.92 | -- | 6.96 |
| Total of change for vegetation types dominated by tall and medium shrubs (degradation) | | | | | 64038 | 38759.76 |
| Total change for all types of natural vegetation | | | | | 233463.88 | 47780.76 |
| A | Intervention areas dominated by anthropogenic deforestation | 56 | A-Hmtin2/Mar, Cul, A-Bhtf2/Sma-Cpl, A-Btin2/Ctr-Ssa, A-BHmhtf2/Mla-Pma and A-Bmhtf2/Xar-Tgu are included | 1120590.6 | 1306273.72 | -- | 185683.12 |

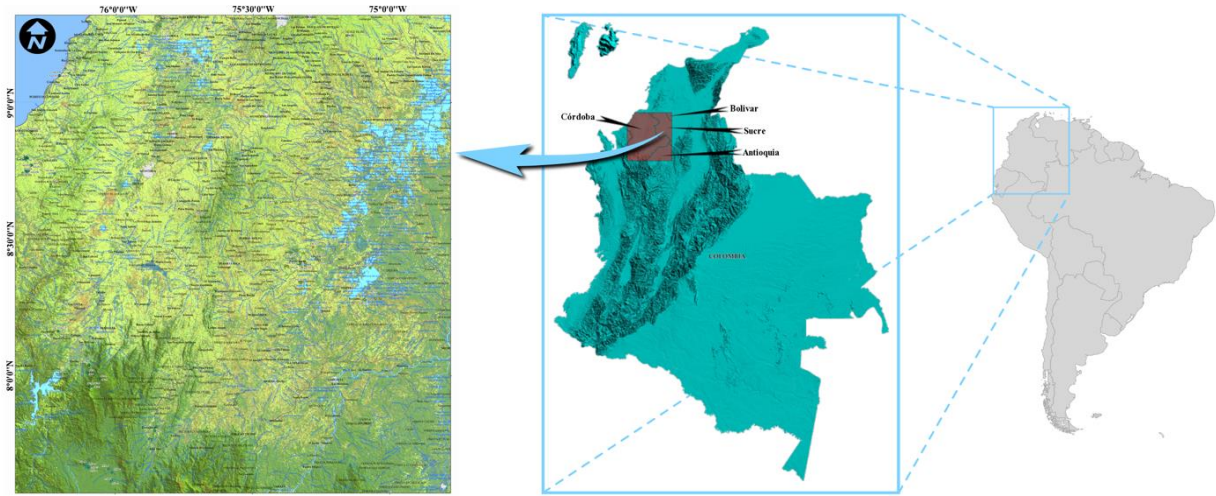

**Fig. 1.** Location of the study area. The Universal Transverse Mercator (UTM) projection of WGS 84 datum zone 18 N (78–72W) was used in all cases.

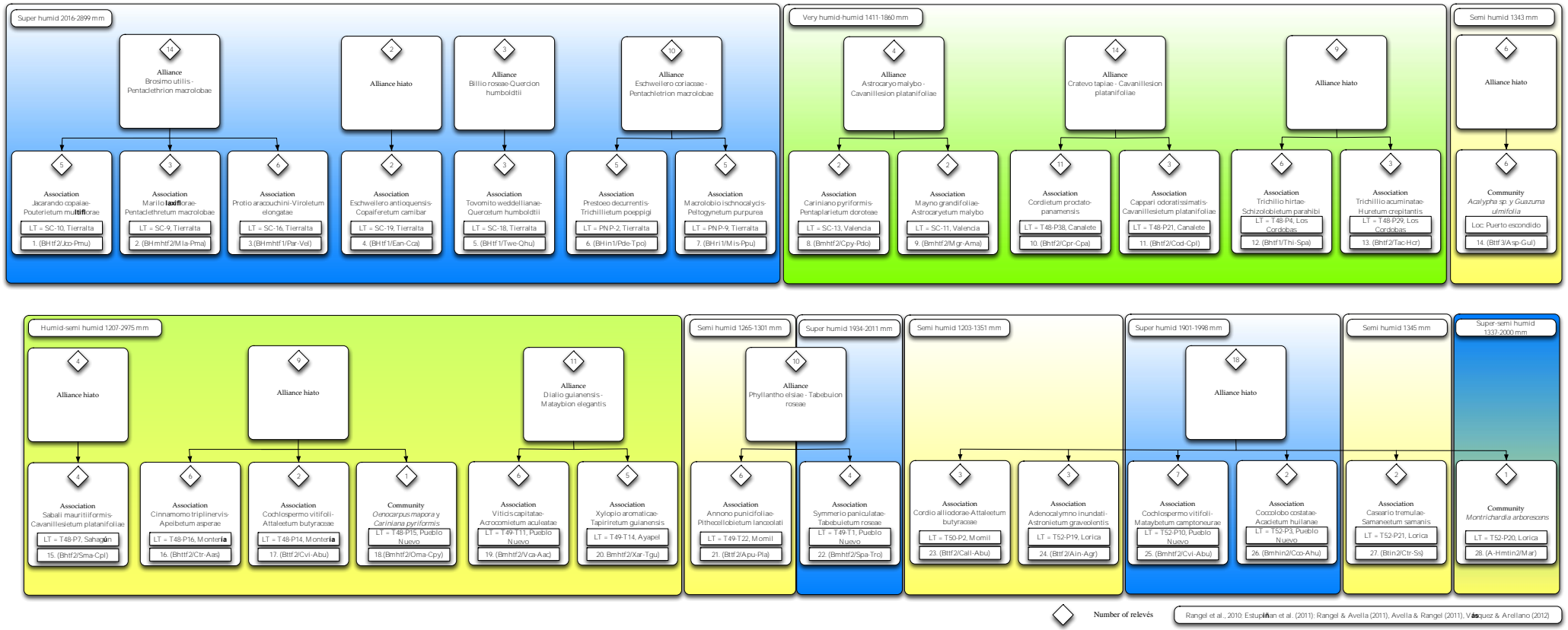

**Fig. 2.** Organization of vegetation types used in the CCA–Fuzzy Land Cover methodology.

**Fig. 3.** Canonical correspondence analysis of species and plots (relevés) for vegetation types in superhumid to semihumid climates (using the physical variable combination of precipitation, altitude, and slope). A. Axes 1 and 2. B. Axes 1 and 3.

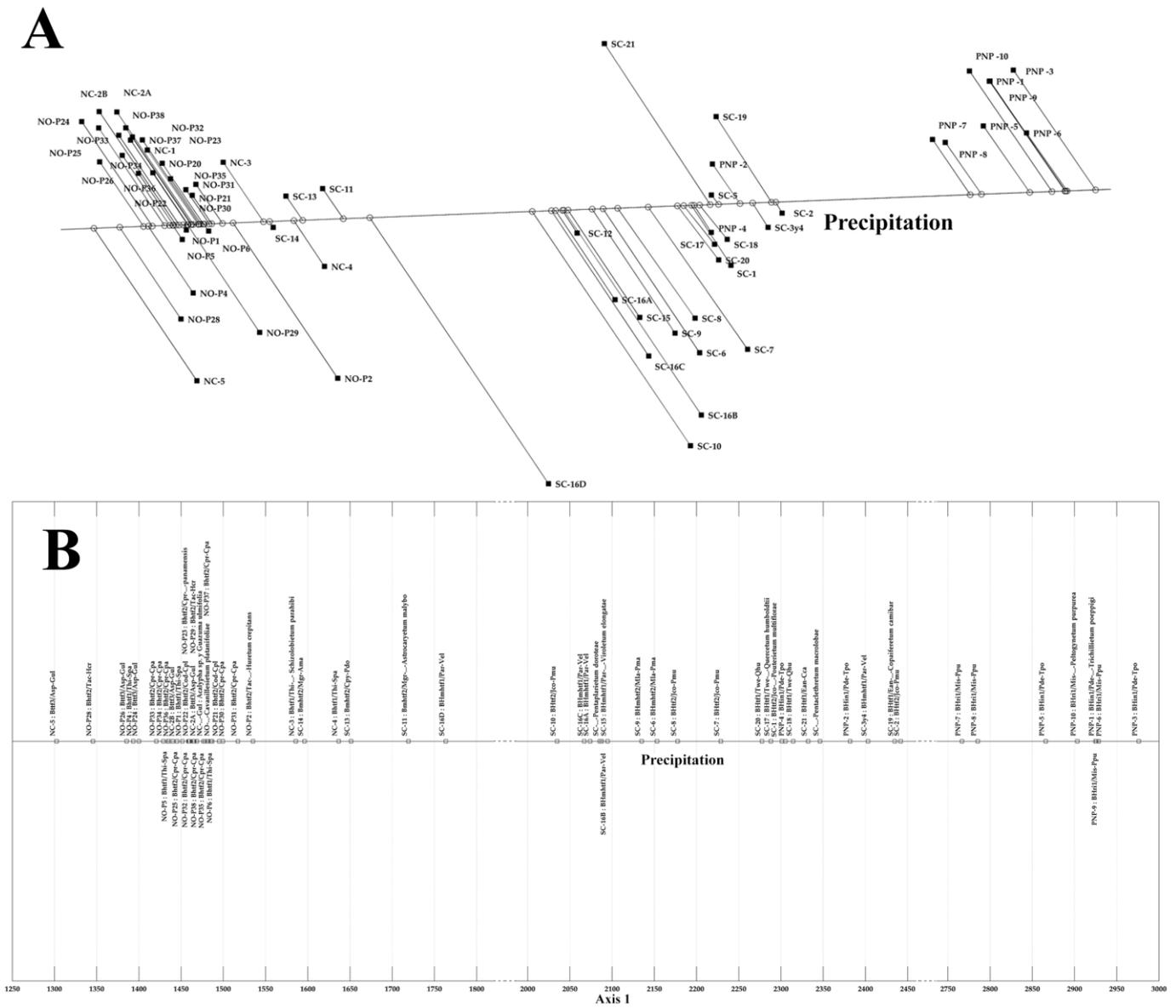

**Fig. 4.** A. Orthogonal projection of plot data for vegetation types in superhumid to semihumid climates along a precipitation gradient. B. Orthogonal projection of plot data calibrated to real precipitation values by linear multiple regression along a precipitation gradient.

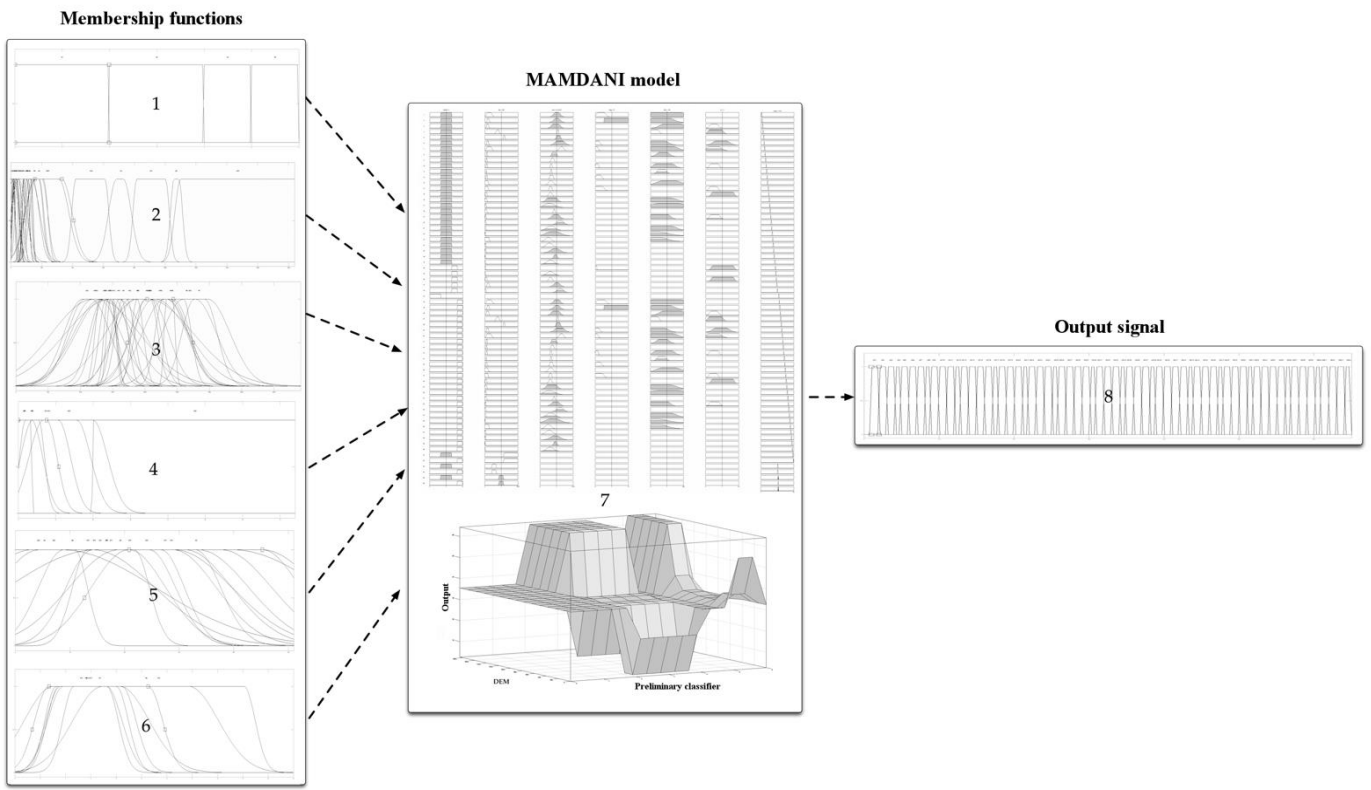

**Fig. 5.** General structure of the fuzzy logic model (data within the membership function), created using MATLAB software for the analyzed decades.

**Fig. 6.** Coverage maps obtained by fuzzy logic modeling, showing anthropic intervention in the study region during the 1980s (A) and during the first decade of the 21st century (B).

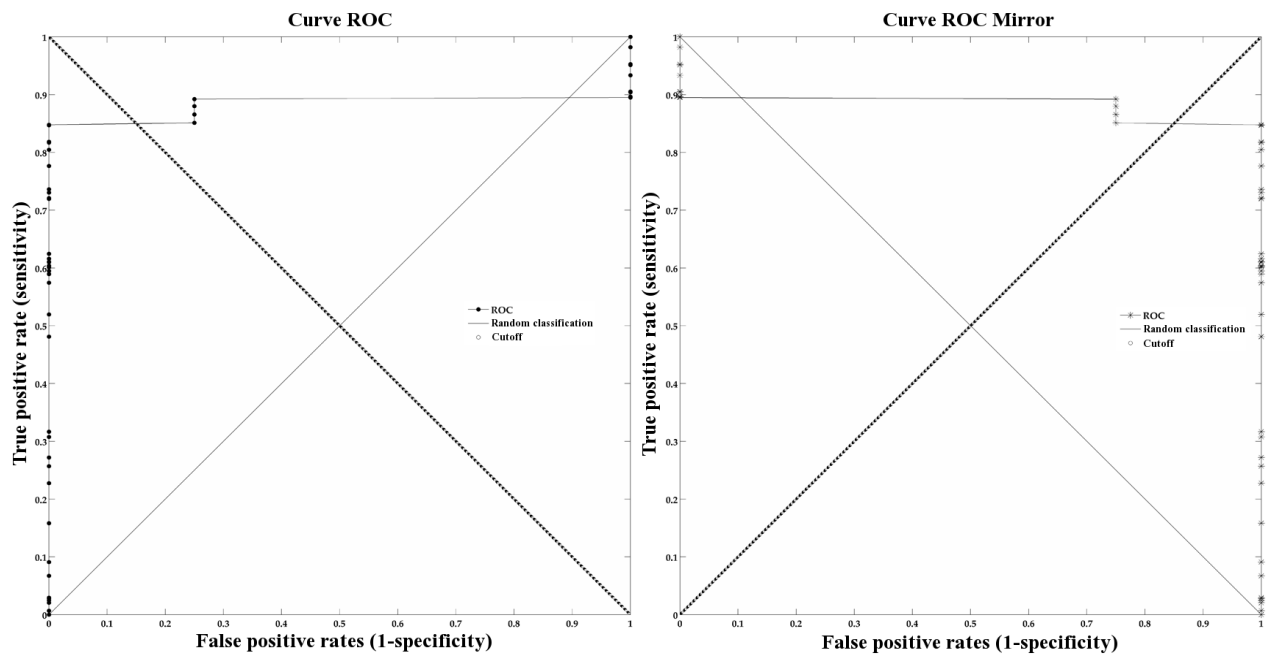

**Fig. 7.** The rates of false positives and true negatives relative to true positive rates corroborate inferences analyzed for each of the cases using Cohen's kappa. According to the area under the curve, the model produces high-quality results (AUC = 0.8827).

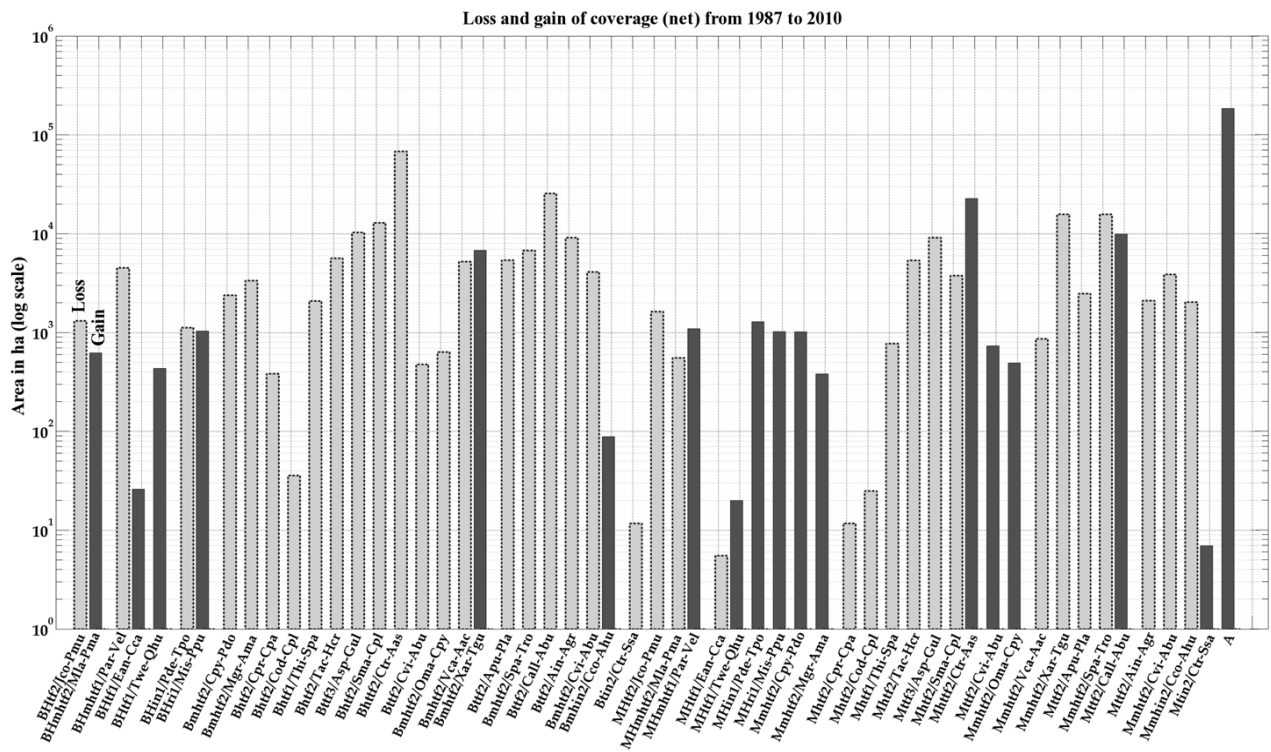

**Fig. 8.** Historical records (1987–2010) of net gain and loss of surface coverages in the study area.